\documentclass[aps,prd,twocolumn,groupedaddress,nofootinbib]{revtex4-2}

\usepackage{amsmath}
\usepackage{amssymb}
\usepackage{graphicx}
\usepackage{float}
\usepackage{multirow}
\usepackage{hyperref}
\usepackage{xcolor}

\begin{document}

\title{Characteristic formulation of the Regge-Wheeler and Zerilli Green functions}

\author{Conor O'Toole}
\author{Adrian Ottewill}
\author{Barry Wardell}
\affiliation{School of Mathematics \& Statistics, University College Dublin, Belfield, Dublin 4, Ireland, D04 V1W8}

\date{\today}

\begin{abstract}

We present a characteristic initial value approach to calculating the Green function of the Regge-Wheeler and Zerilli equations. We combine well-known numerical methods with newly derived initial data to obtain a scheme which can in principle be generalised to any desired order of convergence. We demonstrate the approach with implementations up to sixth-order in the grid spacing. By combining the results of our numerical code with late-time tail expansions and methods of subtracting the direct part of the Green function, we show that the scalar self-force in Schwarzschild spacetime can be computed to better accuracy than previous Green function based approaches. We also demonstrate agreement with frequency domain methods for computing the Green function in the gravitational case.
Finally, we apply the Regge-Wheeler and Zerilli Green functions to the computation of the gravitational energy flux.

\end{abstract}

\maketitle

\section{Introduction\label{sec:Intro}}

While still in its infancy relative to electromagnetic astronomy, gravitational wave astronomy has already seen enormous success in recent years.
The LIGO-VIRGO collaboration has reached the stage of regularly detecting the merger of stellar mass black holes and neutron stars \cite{PhysRevX.9.031040}.
Looking towards future detectors, the planned European Space Agency mission, LISA (Laser Interferometer Space Antenna), will provide access to an entirely new frequency range for the gravitational wave astronomy community \cite{2017arXiv170200786A}.
Operating in the 0.1 -- 100 mHz range, LISA will not only be capable of detecting stellar mass binaries long before merging, but will also detect sources involving much larger masses.
Among the new sources which will be detected for the first time are Extreme Mass Ratio Inspirals (EMRIs).
These systems, comprising a stellar mass compact object such as a black hole, or even a neutron star, orbiting a massive or supermassive black hole ($10^6$ -- $10^9 M_{\odot}$), are expected to be found in the centres of galaxies
and lie in the most sensitive part of the LISA band.
There are a number of expected formation channels for EMRIs, with the most likely being dynamical friction in the dense cluster about the supermassive black hole at a galaxy's centre \cite{AmaroSeoane:2012tx}.
Stars and compact objects within this cluster undergo gravitational interactions which can lead to objects falling into a close orbit around the central black hole.
Main sequence stars are unlikely to survive in this environment due to tidal disruption, but compact objects such as black holes and neutron stars can withstand these tidal forces to produce a long-lived, slowly inspiralling binary.
The expected event rates for detections of EMRIs by LISA remain quite uncertain, being anywhere from 1--$10^3 \ \mathrm{yr}^{-1}$ \cite{Babak:2017tow}.
However, they present some of the most exciting opportunities to study black holes, and also some of the most difficult challenges.

As with LIGO-VIRGO, LISA will identify many of the sources in its frequency band via matched-filtering, comparing detected signals to large banks of precomputed waveforms to pluck out those signals likely to be astrophysical in origin.
This is because most, if not all EMRI signals will actually be so weak as to lie below the noise-level of the detector.
Thus, one of the primary goals of the next decade for the general relativity community is to generate high-accuracy waveforms for all of the expected sources.
Unlike typical LIGO sources, or even other LISA sources such as massive/supermassive black hole binaries, EMRIs spend a large amount of their lifetimes in a region of the parameter space where
many techniques for modelling black hole binaries are not applicable, or not currently practical.
For instance, the orbit of the smaller companion object is typically not circularised, and can be highly inclined.
The system is typically long lived, spending weeks or months within the LISA band.
The smaller object can also reach large velocities, comparable to $c$, during its orbit.
These properties combine to make EMRIs a unique and extremely powerful laboratory in which to test general relativity.
However, they also rule out post-Newtonian theory and numerical relativity as suitable methods to model the system, except in certain limits (eg. at large radii where the companion's velocity is small, where post-Newtonian theory is applicable \cite{2020arXiv200612036V}).
Instead, the standard approach to modelling EMRIs is to compute the self-force (SF) \cite{Poisson:2011nh, Barack:2018yvs, Barack:2009ux}.

\begin{figure}[htb]
    \includegraphics[width = 0.9\columnwidth]{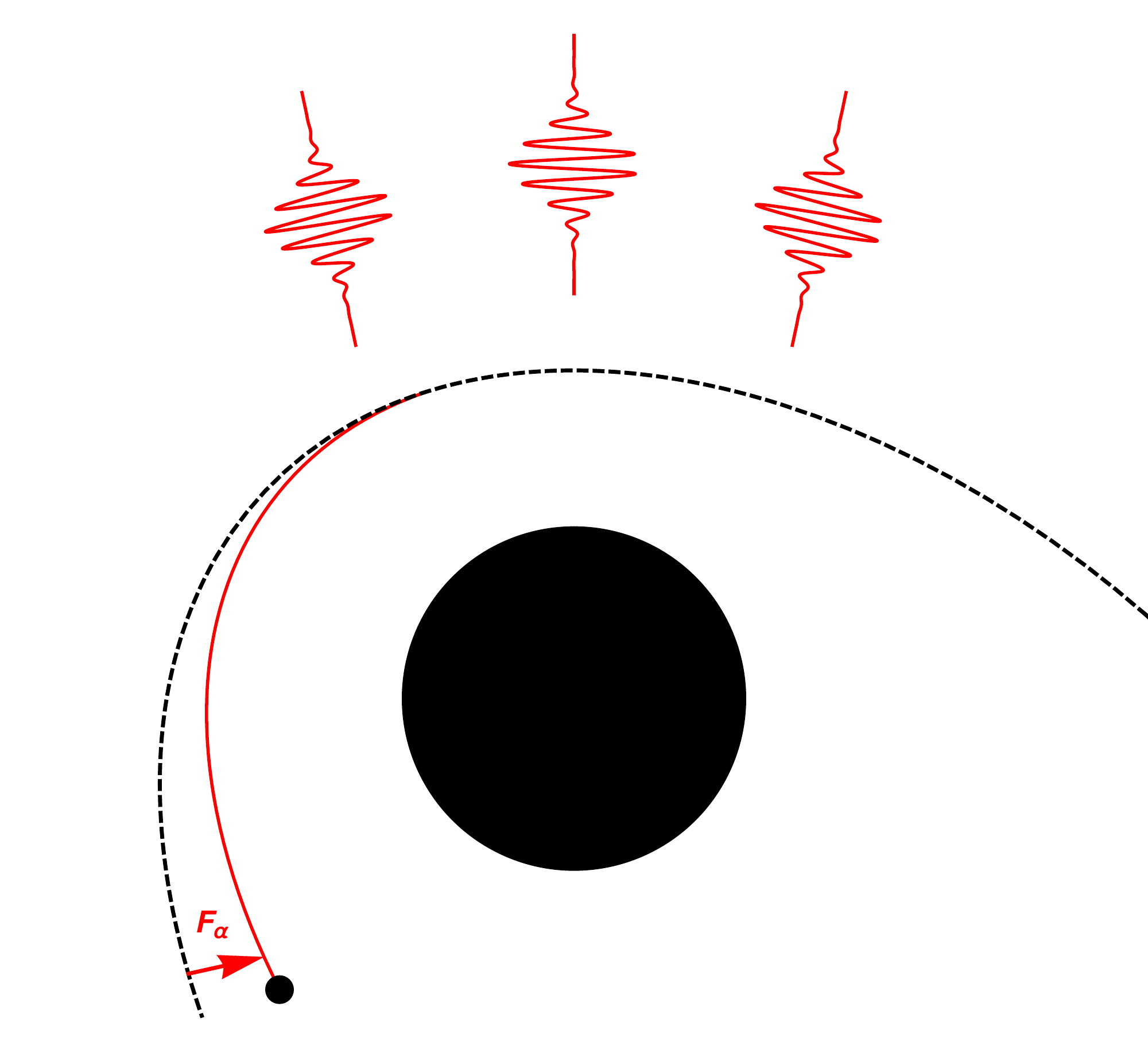}
    \caption[SF explanation]{In the SF picture, the compact object orbiting the central black hole moves along a worldline (red) which deviates from a geodesic of the background spacetime (dashed, black line) due to the emission of gravitational waves. This effect can be described in terms of a (fictitious) force, $F_{\alpha}$ which we call the self-force, as it results from the particle's own bending of spacetime. The effect is exaggerated here, as in reality a background geodesic would be a good approximation to the worldline of the compact object for multiple orbits. However, the magnitude of the SF increases at points on eccentric or hyperbolic orbits which are close to periastron.}
    \label{fig:SFschematic}
\end{figure}

In the SF approach, the effect of the companion (mass $m$) is treated as a small perturbation to the background spacetime of the central black hole (mass $M$).
This leads to an expansion of the metric, order-by-order, in the mass ratio $\epsilon \equiv \frac{m}{M}$,
\begin{equation} \label{eq:MetricPert}
	\mathtt{g}_{\mu \nu} = g_{\mu \nu} + \epsilon h^{(1)}_{\mu \nu} + \epsilon^2 h^{(2)}_{\mu \nu} + \ldots,
\end{equation}
where $\mathtt{g}$ is the exact metric of the full two-body spacetime,
$g$ is the metric of the background, and $h^{(n)}$ is the $n$-th order contribution to the metric perturbation.
Schematically, Fig.~\ref{fig:SFschematic} shows the behaviour of the system.
At zeroth order, the companion would follow a geodesic of the background spacetime.
When higher-order effects are included, the emission of gravitational waves causes the orbit of the companion to slowly evolve and inspiral into the central black hole.

The goal of the self-force approach is to compute the metric perturbation and use it to derive the inspiralling motion of the companion along with the associated gravitational waveform.
There are a number of technical difficulties that arise in the process.
One of the most fundamental is that the metric perturbation is formally singular along the worldline of the companion.
Much of the work in the SF community over the last three decades has been aimed at tackling this problem.
An approach which has seen significant development in recent years is that of \emph{worldline convolution} \cite{Anderson:2004eg,Anderson:2005gb,Casals:2013mpa,Wardell:2014kea}.
This approach makes use of Green functions to construct the metric perturbation by performing convolution integrals over the past worldline of the particle. Once the Green function is available, it is
straightforward to use it to compute the self-force, but this merely moves the challenge to that of computing the Green function.

In this work, we develop a characteristic formulation of the Green function for the Regge-Wheeler and Zerilli equations, which govern perturbations of Schwarzschild spacetime. Similar approaches have previously applied a characteristic method to compute the Green function for the scalar wave equation \cite{Mark:2017dnq,Jonsson:2020npo}. Our approach more fully develops the characteristic formalism, extending these earlier works to be applicable to the Regge-Wheeler and Zerilli equations, and also to arbitrary convergence order in the numerical scheme. The layout of the paper is as follows: in Sec.~\ref{sec:GF} we review the theory of Green functions for the Regge-Wheeler equation; in Sec.~\ref{sec:Num} we describe our numerical method for solving a characteristic initial value problem for the Green function; in Sec.~\ref{sec:ICs} we derive initial conditions that enable the method to be extended to arbitrary convergence order; in Sec.~\ref{sec:Res} we demonstrate our method by applying it to compute the scalar self-force and the gravitational energy flux. Finally, we provide some concluding remarks in Sec.~\ref{sec:Conc}.

Throughout this work we use geometrised units such that the speed of light and the gravitational
constant are set to unity ($G=c=1$).

\section{Green Functions for the Regge-Wheeler Formalism}
\label{sec:GF}
The Regge-Wheeler formalism is based on constructing solutions to the linearised Einstein equations from
solutions to the modified scalar wave equation,
\begin{equation}
    \bigg[\Box + \frac{2Ms^2}{r^3} \bigg] \Psi_s = S_s.
\end{equation}
where $s$ is the spin of the field ($s=0$ for scalar fields, $s=1$ for electromagnetic fields and $s=2$ for gravitational fields).
Working in Schwarzschild coordinates, $\{t,r,\theta,\phi\}$, this equation is separable using
the ansatz
\begin{align}
  \Psi_s &= \sum_{\ell=|s|}^\infty \sum_{m=-\ell}^\ell \, \frac{1}{r} \Psi_{s \ell m}(t, r) \, Y_{\ell m}(\theta, \phi),
\end{align}
where $Y_{\ell m}(\theta, \phi)$ are the spherical harmonics. Transforming to double null coordinates,
$\{u,v\} = \{t-r_\ast, t+r_\ast\}$
where $r_\ast = r+2M \ln (\frac{r}{2M}-1)$ is the Schwarzschild tortoise coordinate, the Regge-Wheeler master function $\Psi_{s \ell m}$ satisfies the Regge-Wheeler equation,
\begin{equation}
  \bigg[ \dfrac{\partial^2}{\partial u \partial v} + \frac{f}{4} \bigg(\frac{\ell(\ell+1)}{r^2} +
  \frac{2M(1-s^2)}{r^3} \bigg)\bigg] \Psi_{s\ell m}  = -\frac{f}{4} S_{s\ell m},
\end{equation}
with $f \equiv 1-\frac{2M}{r}$. Here, the modes of the source are defined in the same way as the master function,
\begin{align}
  S_s &= \sum_{\ell=|s|}^\infty \sum_{m=-\ell}^\ell \, \frac{1}{r} S_{s \ell m}(t, r) \, Y_{\ell m}(\theta, \phi).
\end{align}

We are interested in constructing the retarded Green function, which satisfies
\begin{equation} \label{eq:4DWaveGF}
  \bigg[\Box + \frac{2Ms^2}{r^3} \bigg] G^{\rm ret}_s(x,x') = -4 \pi \delta_{(4)}(x,x'),
\end{equation}
where $\delta_{(4)}(x,x')$ is the invariant Dirac-delta distribution.
Primed coordinates denote the position of the particle, while unprimed coordinates denote a point in the particle's past history.
The retarded Green function can be written in terms of its decomposition into spherical harmonic modes,
\begin{align}
  G^{\rm ret}_s&(x,x') = \nonumber \\
  & \frac{4\pi}{rr'} \sum_{\ell=|s|}^\infty \sum_{m=-\ell}^\ell  G_{s \ell}(r,r'; \Delta t)  Y_{\ell m}(\theta, \phi) Y^*_{\ell m}(\theta', \phi').
\end{align}
Note that time translation invariance means the retarded Green function only depends on the time difference, $\Delta t \equiv t'-t$.
It is convenient to exploit spherical symmetry by using the addition theorem for the spherical harmonics to rewrite the mode decomposition in terms of the angle $\gamma$ between $x$ and $x'$,
\begin{align}
  G^{\rm ret}_s&(x,x') =\nonumber \\
  & \frac{1}{rr'} \sum_{\ell=|s|}^\infty (2\ell+1) P_\ell (\cos \gamma) G^{\rm ret}_{s \ell}(r,r'; \Delta t).
\end{align}
Substituting into Eq.~\eqref{eq:4DWaveGF} and using the completeness relation for the spherical harmonics,
we find that the modes of the retarded Green function satisfy the Regge-Wheeler equation with a distributional source,
\begin{align}
  \bigg[ \dfrac{\partial^2}{\partial u \partial v} + \frac{f}{4} \bigg(\frac{\ell(\ell+1)}{r^2} &+
  \frac{2M(1-s^2)}{r^3} \bigg)\bigg] G^{\rm ret}_{s \ell}(r,r'; \Delta t) \nonumber \\
   &= \frac12 \delta(u'-u) \delta(v'-v).
\end{align}
The modes of the retarded Green function for the Regge-Wheeler equation are thus Green functions for the
flat-space 2D wave equation with a potential. Writing these as
\begin{equation} \label{eq:Gret}
    G^{\rm ret}_{s \ell}(r,r'; \Delta t) = -g_{s\ell}(u,v;u',v') \theta(u'-u) \theta(v'-v)
\end{equation}
we find that $g_{s\ell}(u,v)$ satisfies the homogeneous Regge-Wheeler equation,
\begin{align}
\label{eq:gsl-eq}
  \bigg[ \dfrac{\partial^2}{\partial u \partial v} + \frac{f}{4} \bigg(\frac{\ell(\ell+1)}{r^2} &+
  \frac{2M(1-s^2)}{r^3} \bigg)\bigg] g_{s \ell}(u,v;u',v') = 0,
\end{align}
with characteristic initial conditions
\begin{equation}
\label{eq:gsl-init}
    g_{s\ell} (u, v';u',v') = \frac{1}{2}, \quad g_{s\ell} (u', v;u',v') = \frac{1}{2}.
\end{equation}

\section{Numerical solution of the characteristic initial value problem}
\label{sec:Num}

The characteristic initial value problem represented by Eqs.~\eqref{eq:gsl-eq} and Eqs.~\eqref{eq:gsl-init} is
well-suited to a numerical treatment. All of the numerical methods, as well as the method of deriving initial
conditions, are applicable to any equation of the form of a flat-space wave equation with a potential. As such,
we formulate the scheme for a generic potential, $P(u,v)$, but remind the reader that in the case of the
Regge-Wheeler equation this potential is given by
\begin{equation}\label{eq:Potential}
P^{\mathrm{RW}}(u,v) = -\frac{f}{4} \left( \frac{\ell(\ell+1)}{r^2} + \frac{2 M(1-s^2)}{r^3}\right)
\end{equation}
where $r$ is a function of $v-u$.

In addition, in the gravitational case ($s=2$) we will consider the Zerilli potential,
\begin{equation}\label{eq:ZerPotential}
P^{\mathrm{Zer}}(u,v) = -\frac{f}{4 r^2 \Lambda^2} \left[ 2 \lambda^2 \left( \Lambda + 1 \right) + \frac{18M^2}{r^2}\left( \lambda + \frac{M}{r} \right) \right]
\end{equation}
where $\Lambda = \lambda + \frac{3M}{r}$,
\begin{equation}\label{eq:Lambda}
	\lambda = \frac{1}{2} (\ell+2)(\ell-1).
\end{equation}
The Zerilli equation, Eq.~\eqref{eq:gsl-eq} with the potential replaced by $P^{\mathrm{Zer}}(u,v)$, governs the even
parity perturbations to the Schwarzschild spacetime, while the odd parity perturbations obey the Regge-Wheeler equation with $s=2$.

\begin{figure}[htb]
  \includegraphics[width = 0.9\columnwidth]{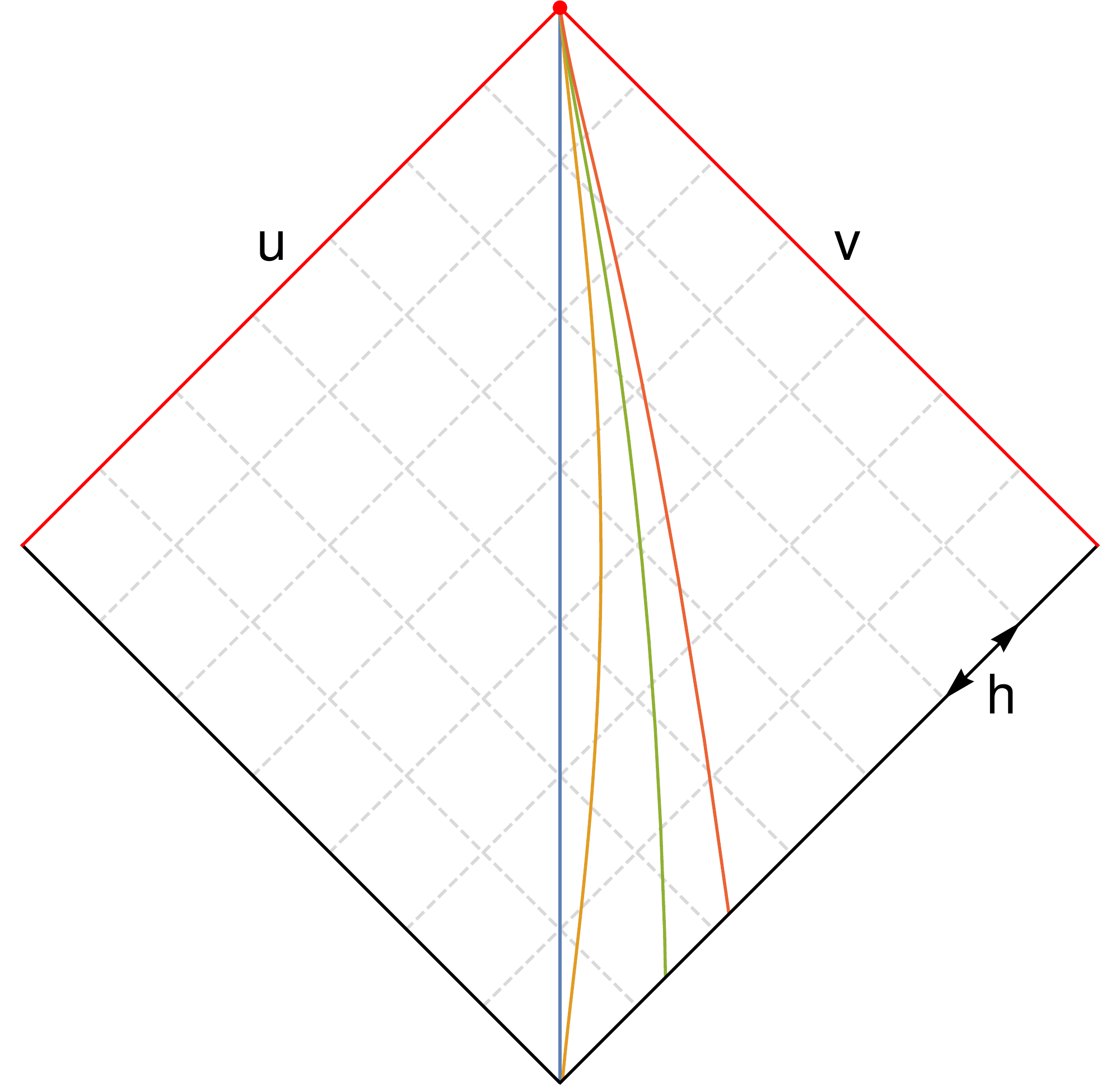}
  \caption[Grid on past light cone]{
    Numerical domain in which we solve for the Green function, with the lightcone in red.
    The red point at the vertex of the light cone is the base point $(u=u', v=v')$. In self-force applications
    this is the location at which the self-field and self-force components for a given worldline are computed.
    The domain contains a large family of past worldlines which pass through this point, several of which are
    shown [projected onto the 2D $(u,v)$ submanifold]. These include a circular orbit (blue), and a number of
    geodesics with increasing eccentricity. Thus for a single run of the numerical code, we can compute the
    self-force at this point for a large number of past worldlines, as opposed to at all points on a given
    worldline as would be done with other methods.
    Note, the horizon is to the left in this plot, while radial infinity is to the right.}
  \label{fig:fullgrid}
\end{figure}

\subsection{Numerical integration stencils}

\begin{figure}[htb]
	\includegraphics[width=0.8\columnwidth]{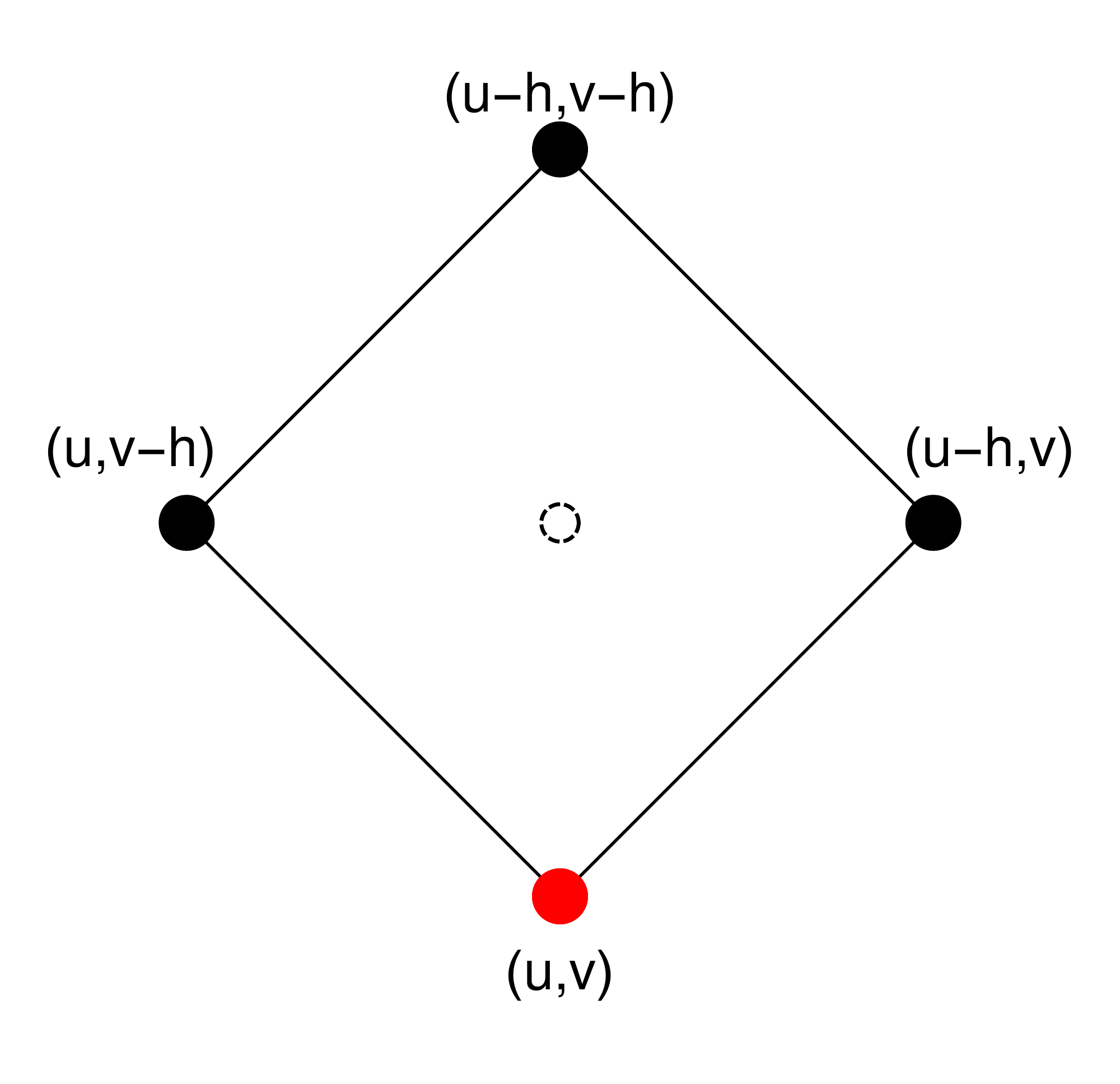}
	\includegraphics[width=0.8\columnwidth]{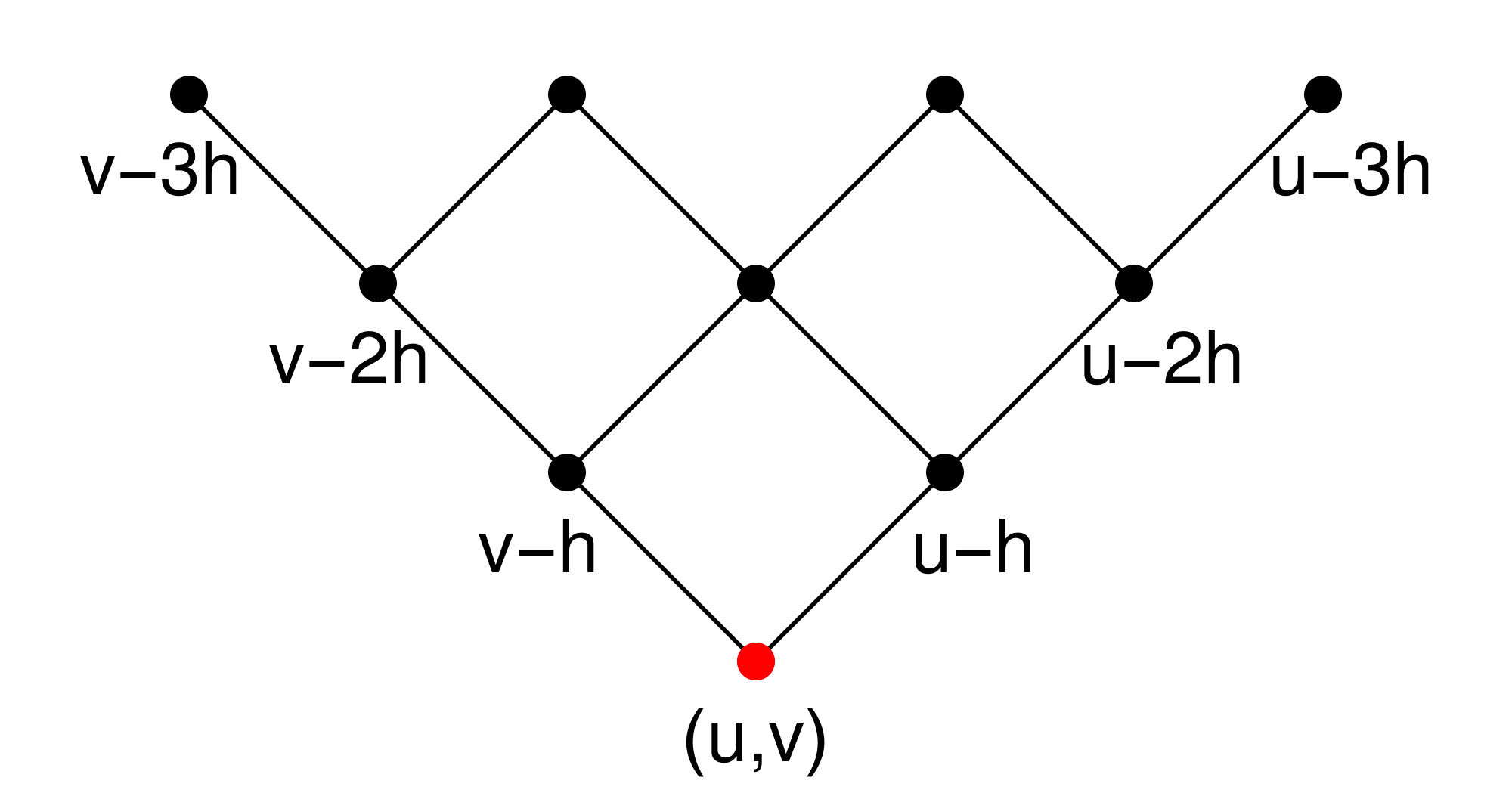}
	\includegraphics[width=0.8\columnwidth]{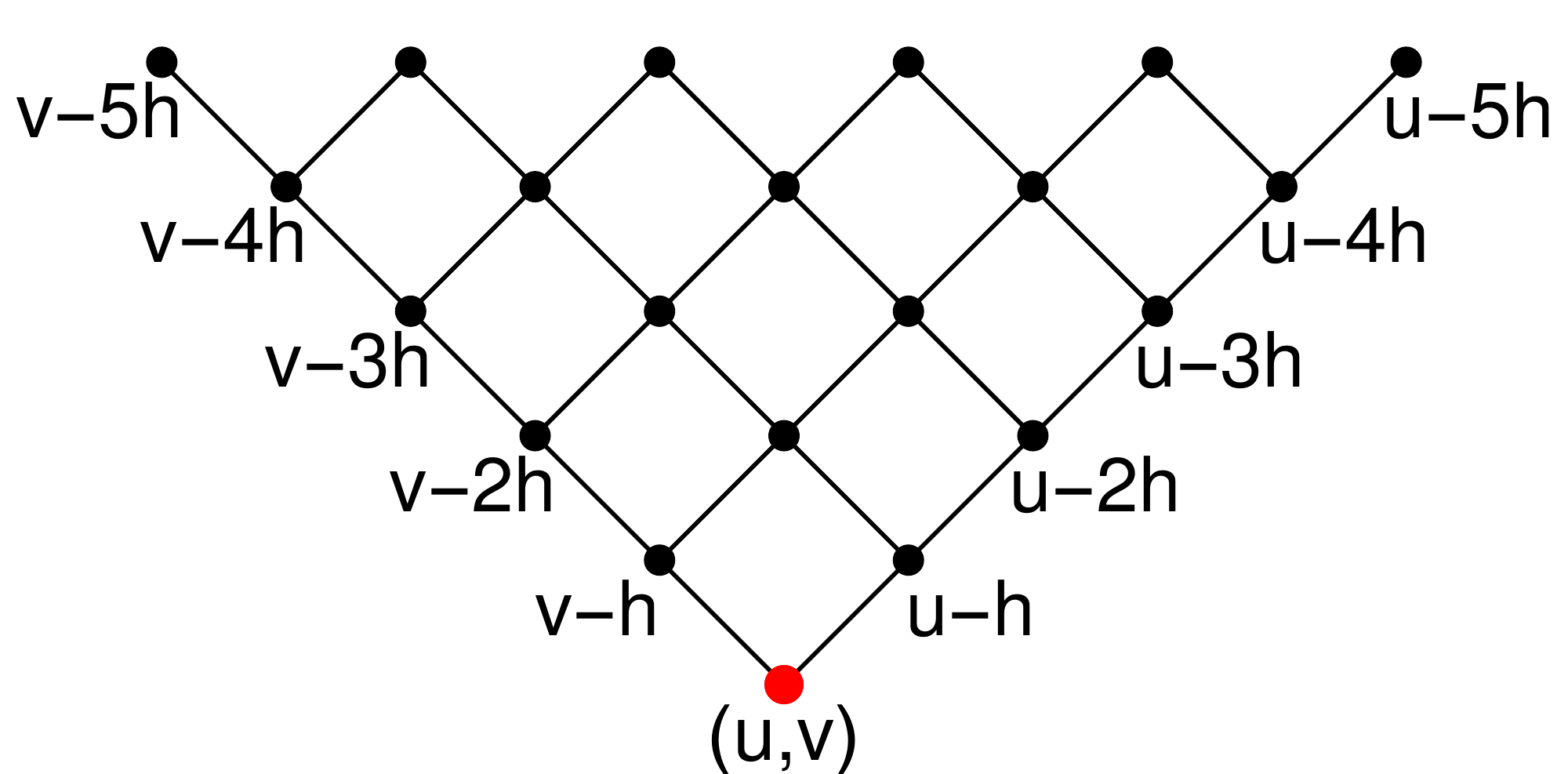}
    \caption[High order stencils]{The stencils for the second order (top), fourth order (middle) and sixth order (bottom) schemes.
        The red point, $(u,v)$, is the point at which we wish to compute the value of $g_{s\ell}$, while it is already known at the other points, indicated in black.
        The hollow point in the centre of the cell in the second order stencil is the point about which the Taylor expansion in Eq.~\eqref{eq:TaylorExp} is performed.
        In the higher order schemes, we show only the coordinate which changes along each ray, for simplicity.
        Note that in each of these diagrams, the lightcone is toward the \emph{top}, so the lower points are further in the past relative to the base point at which we will
        ultimately compute the self-field and self-force components. \label{fig:2nd4th6th}}
\end{figure}

The retarded Green function only has support inside the past light cone, so we need only consider the value of
$g_{s\ell}$ and its derivatives on and inside the lightcone. This suggests a natural manner in which to
subdivide our domain into a grid with spacing $h$, using lines of constant $u$ or $v$, see Fig.~\ref{fig:fullgrid}.
Doing so, we can derive a numerical scheme which will converge as a desired power of $h$.
We begin by considering a single cell, see Fig.~\ref{fig:2nd4th6th}.
Integrating Eq.~\eqref{eq:gsl-eq} over this cell yields (omitting $g$'s dependence on $u'$, $v'$)
\begin{align}
	0 &= \iint \left[ \partial^2_{uv} - P(u,v) \right] g_{s\ell}(u,v) \,{\rm d} u\, {\rm d} v\nonumber \\
	&= \quad g_{s\ell}^{00} + g_{s\ell}^{11} - g_{s\ell}^{10} - g_{s\ell}^{01}  - \iint  P(u,v) g_{s\ell}(u,v) \,{\rm d} u\, {\rm d} v
\end{align}
where we have introduced the notation
\begin{equation}
	g_{s\ell}^{i j} \equiv g_{s\ell}(u-i h, v-j h)
\end{equation}
	to denote the value of the function $g_{s\ell}$ at the point ${(u-i h, v-j h)}$.

Note that the first term in the integral is exact. The second term, however, cannot be integrated exactly and must be evaluated numerically
with an approximation valid to some order in $h$. Using methods from \cite{Barack:2007tm, Barack:2005nr} there is a clear, systematic way to
calculate the integral to any desired order, $\mathcal{O}(h^{n+2})$, where $n$ is even.
With this error in a single cell, we can ensure the global error is $\mathcal{O}(h^n)$ after integrating over $\mathcal{O}(h^{-2})$ intervals.
The fundamental idea behind the method is to Taylor expand the integrand, which we denote by $ H(u,v) = P(u,v) g_{s\ell} (u,v)$ for conciseness,
about the centre point of a given cell,
\begin{equation}\label{eq:TaylorExp}
	H(u,v) = \sum_{i , j = 0}^{\infty} H_{i,j} (u,v) (u-u_c)^i (v-v_c)^j.
\end{equation}
We obtain a factor of $h^2$ from the double integral over the cell, and so we need only take this expansion to $\mathcal{O}(h^n)$, where $n$ is the desired order of global accuracy.
This will result in $\frac{1}{2} n(n +1)$ unknown coefficients $H_{i,j}$, which can be solved for by using the value of the function at known points.

Following this approach, a second order ($n=2$) algorithm is given by
\begin{equation}\label{eq:SecondOrdAlg} \nonumber
 g_{s\ell}^{00(2)} = - g_{s\ell}^{11} + \big(g_{s\ell}^{10} + g_{s\ell}^{01}\big)\big( 1 - \tfrac{h^2}{2} P^{00} \big).
\end{equation}
This scheme has a simple stencil, shown in the top panel of Fig.~\ref{fig:2nd4th6th}.
Assuming we have already solved for, or provided a priori, the values of $g_{s\ell}$ at the points $(u-h, v)$, $(u, v-h)$ and $(u-h, v-h)$, we can thus
compute the value of $g_{s\ell}$ at the point $(u,v)$ to $\mathcal{O}(h^4)$.

The overall numerical implementation of this algorithm is then straightforward.
We provide initial data along the lightcone (described in more detail in Sec.~\ref{sec:ICs}), and evolve through the domain.
The most straightforward evolution method is to evolve down each ray of fixed $u$ or $v$, then move on to the next, $u+h$ or $v+h$.
We have implemented this algorithm in a C code, with two rays stored in memory at any one time, and specific rays saved for later output.
This allows for the code to use comparatively little memory, even for large domains and fine resolutions.

Higher order schemes are straightforward to obtain using the same method.
We have taken this to sixth order as a demonstration, with the algorithms derived using the computer algebra software Mathematica.
The only obstacle to even higher order schemes is the computational cost of the derivation, though it is not expected that anything beyond eighth order would prove necessary for most practical applications.
A fourth order algorithm is given by
\begin{widetext}
\begin{align}
&g_{s\ell}^{00(4)}  = -g_{s\ell}^{11} + g_{s\ell}^{01} + g_{s\ell}^{10} \nonumber \\
   			 & \quad - \tfrac{h^2}{24} \Big[ 2 P^{00} g_{s\ell}^{00(2)} + 10 \big( P^{11} g_{s\ell}^{11} + P^{01} g_{s\ell}^{01} + P^{10} g_{s\ell}^{10}\big)  - 4 \big(P^{20} g_{s\ell}^{20} + P^{02} g_{s\ell}^{02}\big)+ \big(P^{30} g_{s\ell}^{30} + P^{03} g_{s\ell}^{03} - P^{12} g_{s\ell}^{12} - P^{21} g_{s\ell}^{21}\big)\Big]
		\label{eq:FourthOrdAlg}
\end{align}
while a sixth order algorithm is given by
\begin{align}
g_{s\ell}^{00(6)} & = -g_{s\ell}^{11} + g_{s\ell}^{01} + g_{s\ell}^{10} - \tfrac{h^2}{1440} \Big[108 P^{00} g_{{s\ell}}^{00 (4)} + 371\big(P^{03} g_{s\ell}^{03} + P^{30}g_{s\ell}^{30}\big)  - 154 \big(P^{40}g_{s\ell}^{40} + P^{04}g_{s\ell}^{04}\big) \nonumber \\
  & + 116 \big(P^{31} g_{s\ell}^{31} + P^{13}g_{s\ell}^{13}\big) + 40 P^{22} g_{s\ell}^{22} + 27 \big(P^{50} g_{s\ell}^{50} + P^{05} g_{s\ell}^{05}\big) - 19 \big(P^{41} g_{s\ell}^{41} + P^{14} g_{s\ell}^{14}\big)- 5 \big(P^{32} g_{s\ell}^{32} + P^{23} g_{s\ell}^{23}\big) \nonumber \\
  & + 627 \big(P^{10} g_{s\ell}^{10} + P^{01} g_{s\ell}^{01}\big) + 1032 P^{11} g_{s\ell}^{11} - 504 \big(P^{20} g_{s\ell}^{20} + P^{02} g_{s\ell}^{02}\big)  - 329 \big(P^{21} g_{s\ell}^{21} + P^{12} g_{s\ell}^{12}\big)\Big]
  \label{eq:SixthOrdAlg}
\end{align}
\end{widetext}
The stencils for these schemes are shown in the lower panels of Fig.~\ref{fig:2nd4th6th}.

Note that in the fourth order scheme, the second order approximation appears.
This is because the point at which we wish to calculate the value of the Green function is being used in the derivation of the algorithm.
As we do not yet know the value at this point, we must provide it to a suitable approximation.
Given that it is multiplied by an overall factor of $h^2$, we can use the second order scheme, and the overall error will remain $\mathcal{O}(h^6)$.
We use a similar approach for the sixth order scheme, with the fourth order approximation used in the algorithm.

Our algorithms are by no means unique.
A notable alternative is the predictor-corrector method used in \cite{Barack:2007tm}.
However, the Lorenz gauge equations solved there involve first derivatives, which reduce the order by the factor of $h$.
We can thus avoid employing such a scheme, which would impact numerical efficiency due to being an iterative method, as we do not have any such first order
derivatives in the Regge-Wheeler or Zerilli equations.

As is clear from Fig.~\ref{fig:2nd4th6th}, as we go to higher order we require a greater amount of past information in order
to calculate the value of the Green function at a given point.
This has significant implications for the initial data which must be provided, as will be discussed in detail in Sec.~\ref{sec:ICs}.
Aside from initial data issues, the evolution through the numerical domain can be implemented in the same manner as for the second order scheme, though requiring more rays to be stored in memory during the computation.
However, this scales only as $n$, whereas the scheme is accurate to $\mathcal{O}(h^n)$.
Thus moving to higher order schemes, while computationally slower due to the increased number of operations, is not a significant additional burden on memory resources.

\subsection{Calculation of derivatives of the Green function}

In addition to solving for the Green function, we can also use the method outlined above to solve directly for the derivatives of the Green function, as opposed to computing such derivatives in post-processing by means of finite difference.
For the first derivative with respect to the base point (the vertex of the lightcone),which is required for computing the self force, we can simply apply the above algorithms to equations for the derivatives,
\begin{subequations}
\begin{equation}
\label{eq:UpDeriv}
	\big[\partial^2_{uv} - P(u,v)\big]\partial_{u'}g_{s\ell}(u,v) = 0
\end{equation}
\begin{equation}
\label{eq:VpDeriv}
	\big[\partial^2_{uv} - P(u,v)\big]\partial_{v'}g_{s\ell}(u,v) = 0
\end{equation}
\end{subequations}

Higher order derivatives with respect to the base point are equally straightforward.
However, we can also compute derivatives with respect to the field point (a point inside the past worldline), as well as mixed derivatives with respect to both the base point and field point.
Applying the desired differential operator to the generalised equation,
\begin{subequations}
	\begin{equation}
	\label{eq:UDeriv}
		\big[\partial^2_{uv} - P(u,v)\big]\partial_{u}g_{s\ell}(u,v) - \partial_{u}P(u,v)g_{s\ell}(u,v) = 0
	\end{equation}
	\begin{equation}
	\label{eq:VDeriv}
		\big[\partial^2_{uv} - P(u,v)\big]\partial_{v}g_{s\ell}(u,v) - \partial_{v}P(u,v)g_{s\ell}(u,v) = 0
	\end{equation}
\begin{equation}
\label{eq:UUpDeriv}
	\big[\partial^2_{uv} - P(u,v)\big]\partial_{uu'}g_{s\ell}(u,v) - \partial_{u}P(u,v)\partial_{u'}g_{s\ell}(u,v) = 0
\end{equation}
\begin{equation}
\label{eq:UVpDeriv}
	\big[\partial^2_{uv} - P(u,v)\big]\partial_{uv'}g_{s\ell}(u,v) - \partial_{u}P(u,v)\partial_{v'}g_{s\ell}(u,v) = 0
\end{equation}
\end{subequations}
for example, we have equations to which our schemes may be applied.
$g_{s\ell}(u,v)$ and lower order derivatives that appear can be computed alongside the desired function.
Further combinations of base point and field point derivatives, as well as higher order derivatives can be obtained in a similar fashion.

\subsection{Convergence}

\begin{figure}[htb]
    \includegraphics[width = \columnwidth]{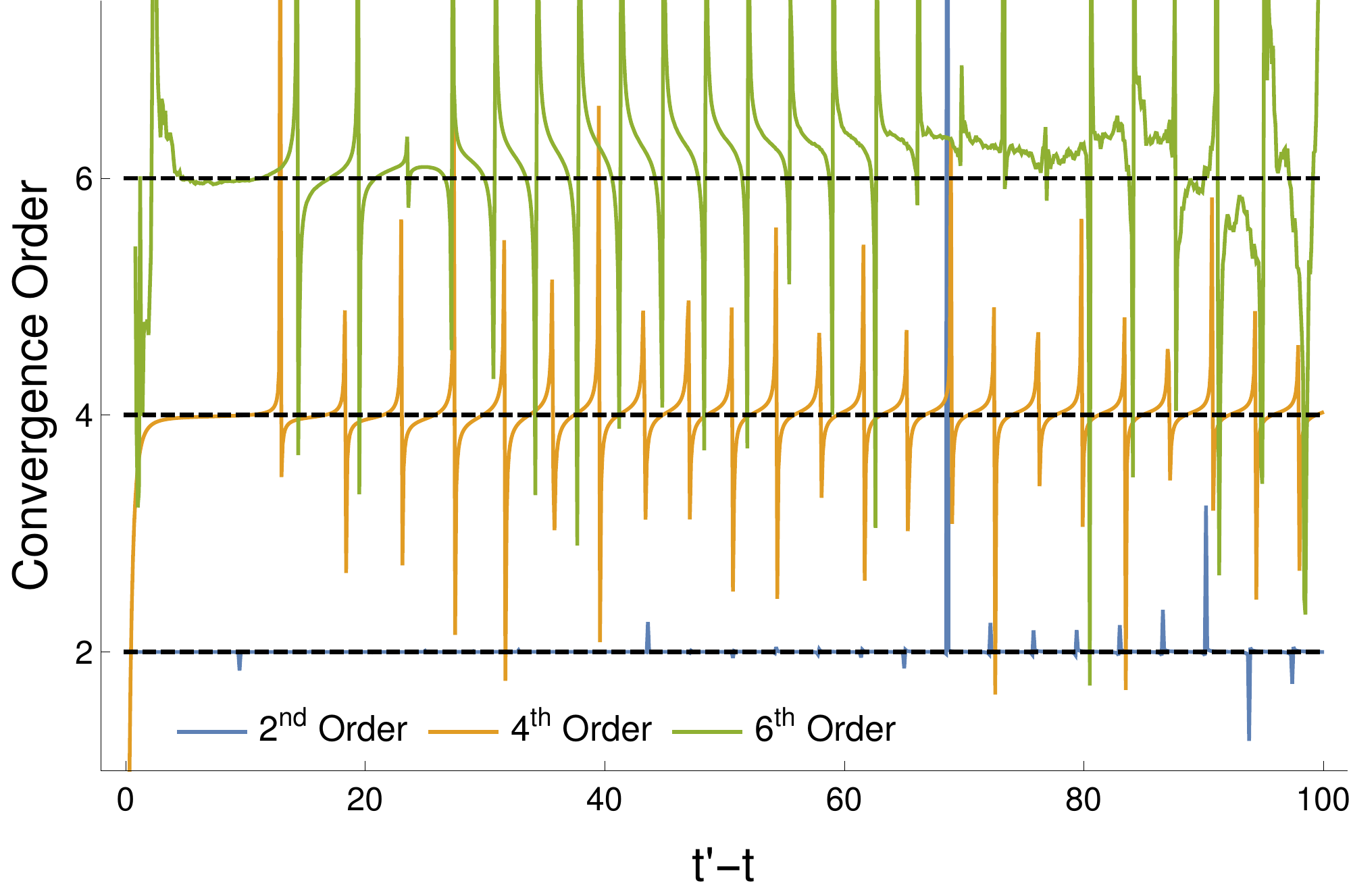}
    \caption[Convergence order for the schemes.]{Convergence order of the numerical calculation of the Regge-Wheeler, $s=0$, Green function along $r=r'$ for $\ell=4$.
        The black, dashed lines indicate the expected convergence orders for the schemes.
        The spikes in the convergence order are associated with zero-crossings, and are not unusual in numerical methods.
        Note, in addition, the noise in the sixth order scheme.
        This is due to roundoff error as we go to higher resolutions.\label{fig:ConvOrd}}
\end{figure}

The schemes outlined here have been implemented in a C code, up to sixth order for the Regge-Wheeler Green function, and fourth order for its derivatives.
The Zerilli Green function and its derivatives have been taken to second order, though extending these to higher orders is straightforward.
The convergence order of these schemes (once sufficiently accurate initial data is provided) can be verified by comparing numerical solutions
using three different resolutions: $h_L$, $h_M$ and $h_H$, which we denote by $g_{s\ell}^{(L)}$,
$g_{s\ell}^{(M)}$ and $g_{s\ell}^{(H)}$, respectively. If the scheme converges as $\mathcal{O}(h^n)$, then $n$ can be calculated by solving
\begin{equation}
\label{eq:ConvOrder}
\frac{g_{s\ell}^{(H)} - g_{s\ell}^{(M)}}{g_{s\ell}^{(M)} - g_{s\ell}^{(L)}} = \frac{h_{H}^n - h_{M}^n}{h_{M}^n - h_{L}^n}.
\end{equation}
The value of $n$ for a single $\ell$ mode of the Regge-Wheeler Green function computed using each of the three numerical algorithms is shown in Fig.~\ref{fig:ConvOrd}, and verifies
that they do indeed converge at the expected order.

\subsection{Code optimisation considerations}

Two additional steps have been taken to improve the efficiency and computation time of the code.
\begin{enumerate}
  \item OpenMP parallelization has been implemented, with independent $\ell$--modes computed in parallel.
  \item Given that the potential, $P$ and $f$ are effectively functions of $v-u$, not general functions of $(u,v)$, they need only be computed at points on the lightcone, as a vertical
        line through the domain shown in Fig.~\ref{fig:fullgrid} marks a line of constant $r$. This also holds for the computation of $r$ from $r_\ast$, which is
        performed numerically and is the most expensive individual pointwise calculation performed in the code. Thus, significant gains in computational cost can
        be obtained by precomputing these before evolution through the domain.
\end{enumerate}
The current implementation of our code can compute 101 modes, using the fourth order scheme with $h=10^{-2}$, up to $\Delta t=240$ in the past of the base point, and parallelized over 5 threads in
217 s, on a desktop computer with a Ryzen 7 processor.
Scripts to automate the calculation over a range of base points make it possible to compute the Green function in a large region outside a Schwarzschild black hole, though this does
lead to a substantial data storage problem.
We expect the use of reduced order surrogate models to provide a solution to this in the future.

\section{Initial Data\label{sec:ICs}}

With the numerical scheme outlined, it remains to provide suitable initial data to compute the retarded Green function.
Previous implementations of similar numerical schemes have not been concerned with initial data, as they chose to allow junk radiation sourced by inconsistent initial data to radiate away, and only consider the solution far from the lightcone where the initial data is imposed \cite{Barack:2007tm, Barack:2005nr}.
However, for the purposes of computing the Green function using a characteristic initial value formulation, the early time behaviour near the the light cone is crucial to obtaining the correct Green function.
As discussed in Sec.~\ref{sec:Num}, the higher the order of the scheme, the greater the number of values of $u$ and $v$ at which the value of the Green function must be known a priori.
As a result, for higher orders, we must provide initial data on rays inside the lightcone, not just on the lightcone itself.

Providing values on the lightcone alone would only be sufficient for the second order scheme.
The value of $g_{s\ell}(u,v; u', v') $ along null rays connecting $(u,v)$ and $(u',v')$ is known to be $\frac{1}{2}$,
and this alone provides sufficient initial data for the second order scheme.

For the higher order schemes, rather than solve for the value of the Green function on points inside the lightcone exactly, we use an expansion of $g_{s\ell}$ near the lightcone to the same order in $h$ as our numerical scheme. Writing
\begin{equation}\label{eq:Ansatz}
	g_{s\ell}(u, v; u', v') \approx \sum_{k = 0}^{n} V_k (u, v; u', v') \left( -\tfrac{\Delta u \Delta v}{2} \right)^k,
\end{equation}
where $\Delta u = u-u'$ scales as $h$ for rays near the $\Delta u=0$ side of the lightcone and $\Delta v$ scales as $h$ for rays near the $\Delta v=0$ side of the lightcone, this approximation is accurate to $\mathcal{O}(h^n)$\footnote{Near the the vertex of the lightcone, both $\Delta u$ and $\Delta v$ scale as $h$ so the approximation is even better and is accurate to $\mathcal{O}(h^{2n})$.}.
Since the initial data is only required on a small, $h$-independent number of rays near the lightcone, this would then be sufficient to ensure that the
global error remains $\mathcal{O}(h^n)$.

We can derive analytic expressions for the $V_k$'s for the both Regge-Wheeler and Zerilli functions by substituting the above expansion into the relevant equation and solving the transport equations for each coefficient, order by order.
Considering the equation with a general potential and demanding that it be satisfied at each power of $\tfrac{\Delta u \Delta v}{2}$ we obtain the transport equations
\begin{align}\label{eq:FullTrans}
(k+1)(\Delta u \partial_u + \Delta v \partial_v + (k+1)) V_{k+1} = -2 \big(\partial^2_{uv} - P \big)V_k
\end{align}
along with the initial condition $V_0 = \frac{1}{2}$ for the value of the Green function on the lightcone.

The equation for $V_1$ can be solved directly using the method of characteristics to obtain an integral solution,
\begin{equation} \label{eq:V1int}
V_1 = \frac{1}{\Delta u} \int_{u'}^u P\left(k, \frac{(k-u')\Delta v}{\Delta u} +v' \right) dk.
\end{equation}
In the case of the Regge-Wheeler and Zerilli equations where $P(u,v)$ is in reality a function of a single variable, $r$,
this result simplifies and we can evaluate the integral explicitly to get
\begin{widetext}
\begin{subequations}
	\begin{equation}
		V^{\mathrm{RW}}_1(u,v;u',v') = \frac{(r-r')(2(\lambda+1) r r' - M (s^2-1) (r + r'))}{4 r^2 r'^2 \Delta r_\ast}
	\end{equation}
	\begin{equation}
		V^{\mathrm{Zer}}_1(u,v;u',v') = \frac{(r-r') \left(9 M^3 (r+r')+3 \lambda  M^2 \left(r^2+4 r r'+r'^2\right)+3 \lambda ^2 M r r' (r+r')+2
   \lambda ^2 (\lambda +1) r^2 r'^2\right)}{2 r^2 r'^2 (3 M+\lambda  r) (3 M+\lambda  r')  \Delta r_\ast}.
	\end{equation}
\end{subequations}

The equation for $V_2$ can be solved in a similar fashion, and its solution is given by

\begin{subequations}
\begin{align}
  V^{\mathrm{RW}}_2 &= \frac{(r-r') (2(\lambda+1) r r'-M (s^2-1) (r+r'))}{8 r^2 r'^2 (\Delta r_\ast)^3} \nonumber \\
    & \quad +\frac{1}{32 r^4 r'^4 (\Delta r_\ast)^2} \{ 4(\lambda+1) r^2r'^2 [ \lambda(r-r')^2 -2r r' ] +M^2(s^2-1)[(s^2-9)(r^4+r'^4)-2r^2r'^2(s^2-1)] \nonumber \\
		& \quad - 4Mr r'(r+r')[(r^2+r'^2)(\lambda(s^2-3)-2) -r r'(2\lambda(s^2-2)+s^2-3)]
		 \}.
\end{align}
\begin{align}
  V^{\mathrm{Zer}}_2 &= \frac{(r-r') \left(9 M^3 (r+r')+3 \lambda  M^2 \left(r^2+4 r r'+r'^2\right)+3 \lambda ^2 M r r' (r+r')+2
   \lambda ^2 (\lambda +1) r^2 r'^2\right)}{4 r^2 r'^2 (3 M+\lambda  r) (3 M+\lambda r') (\Delta r_\ast)^3} \\ \nonumber
	 & \quad +\frac{1}{16 r^4 r'^4 (3 M+\lambda  r) (3 M+\lambda  r') (\Delta r_\ast)^2} \{9 M^4
   \big(9 r^4-2 r^2 r'^2+9 r'^4\big)+9 M^3 \big[3 \lambda r^5+(7 \lambda -4) r^4 r'\\ \nonumber
	 & \quad -2 \lambda  r^3 r'^2-2 \lambda
    r^2 r'^3+(7 \lambda -4) r r'^4+3 \lambda  r'^5\big]+3 \lambda  M^2 r r' \big[(7 \lambda -4) r^4+4 (\lambda -2)
   r^3 r'-6 \lambda  r^2 r'^2 \\ \nonumber
	 & \quad +4 (\lambda -2) r r'^3+(7 \lambda -4) r'^4\big]+4 \lambda ^2 (2 \lambda -1) M r^2
   r'^2 (r^3+r'^3)+4 \lambda ^2 (\lambda +1) r^3 r'^3 \big[\lambda  r^2-2 (\lambda +1) r r'+\lambda r'^2\big] \}.
\end{align}
\end{subequations}
\end{widetext}
For higher order terms, we employ an ansatz for each term,
\begin{equation} \label{eq:VnAnsatz}
	V_k = \sum_{\substack{m=k, \\ m \text{ even}}}^{2k - 1} \frac{p_{km}(r,r')}{(\Delta r_\ast)^m} + \sum_{\substack{m=k,\\ m \text{ odd}}}^{2k - 1} p_{km}(r,r') \frac{(r-r')}{(\Delta r_\ast)^m},
\end{equation}
and solve the resulting equations for the coefficients $p_{km}$ using Mathematica.
We applied this procedure to compute up to $V^{\mathrm{RW}}_8$ and $V^{\mathrm{Zer}}_6$.
We do not give the higher order coefficients here due to the length of their expressions, but have provided them electronically as supplemental material.
Although deriving further $V_k$'s is trivial, we chose to stop at $V^{\mathrm{RW}}_8$ as that is already more than is required for our sixth-order numerical algorithm.

Note that there is no $t$ dependence in $V^{\mathrm{RW/Zer}}_1$ or $V^{\mathrm{RW/Zer}}_2$.
In fact, it is trivial to show, by induction, that $V^{\mathrm{RW/Zer}}_k$ depends only on $r$ for all $k$;
the $t$ dependence in the initial conditions is contained entirely within the powers of $-\frac{\Delta u \Delta v}{2}$.
This is what allowed for the transport equations to be simplified to the point that they could be solved analytically.

\subsection{Initial data near the vertex of the lightcone}

\begin{figure}[htb]
    \includegraphics[width = \columnwidth]{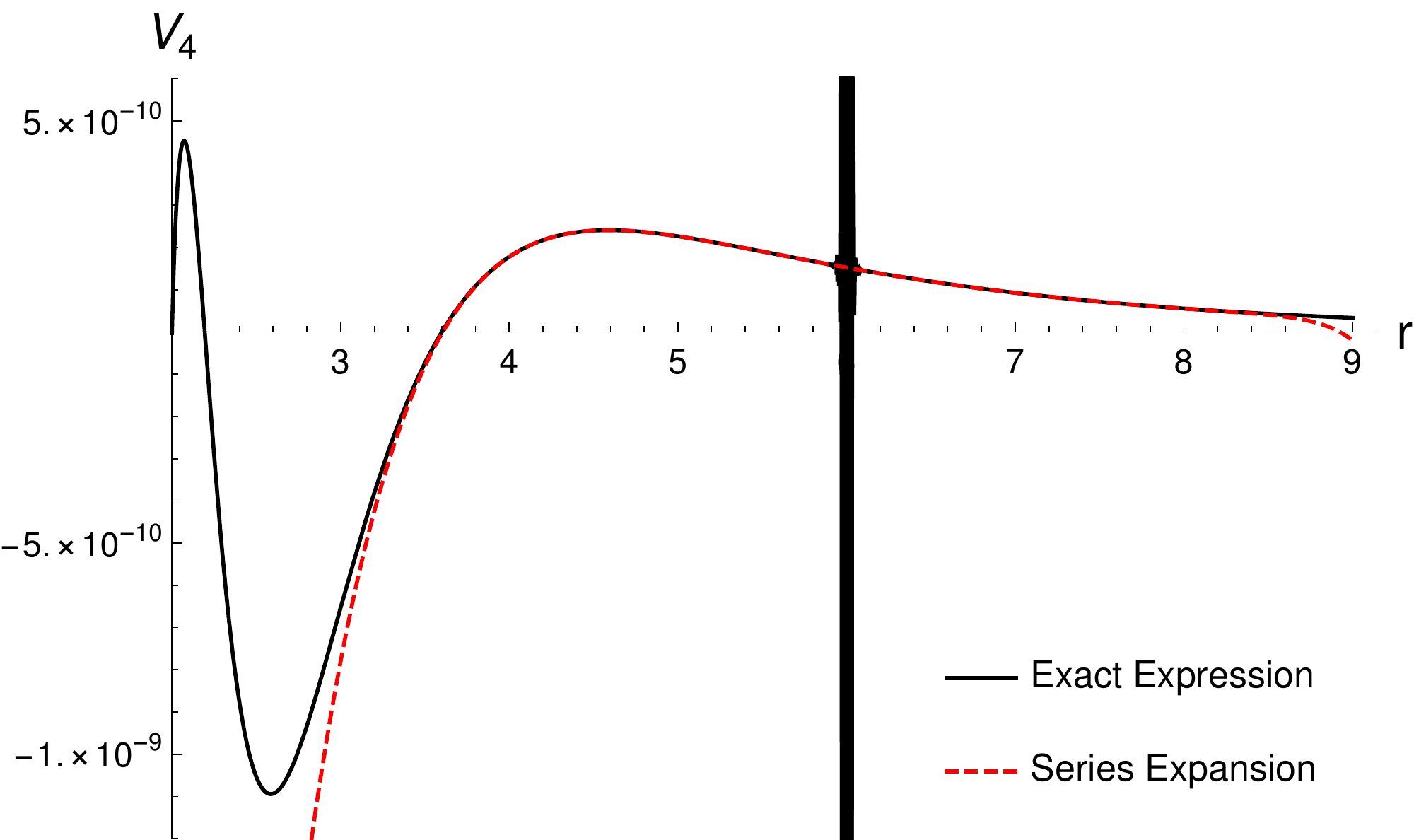}
    \caption[Plot of the function V4.]{Plot of $V^{\mathrm{RW}}_4$ versus radius, for the case of the vertex of the lightcone located at $r'=6$.
        The behaviour at $r=r'$ can be avoided by use of the series expansion.\label{fig:V4Plot}}
\end{figure}

The expressions for the $V^{\mathrm{RW/Zer}}_k$'s appear to be singular at coincidence, where the denominator $\Delta r_\ast$ vanishes.
However, non-singularity is explicitly enforced when solving for them so we know that the apparent singularity will be cancelled by the numerators also vanishing.
This cancellation is not explicit due to the non-trivial relationship between $r$ and $r_\ast$.
In order to avoid this becoming an issue in the numerical implementation, where cancellation is only every guaranteed to the level of round-off, we employ a series expansion about coincidence for points very near the lightcone. Figure \ref{fig:V4Plot} shows the effect of this numerical cancellation and its resolution
via our use of a series expansion.

\subsection{Initial data for the derivatives of the Green function}

By differentiating the functions $V_k$, we can obtain initial data for the derivatives of the Green function.
This requires deriving the coefficients for one order higher than for the Green function, but there is no significant challenge to this.
For instance, for the first $u'$ derivative, say, we have the expression
\begin{align}
	\label{eq:UDerivICs}
	\partial_{u'} g_{s\ell}= \sum_{k = 1}^{\infty} \big(  \tfrac{k \Delta v}{2} V_k -\tfrac{\Delta u \Delta v}{2} \partial_{u'} V_k
       \big)\big( -\tfrac{\Delta u \Delta v}{2} \big)^{k-1}.
\end{align}
Having lost an order in $h$ in the first term in the sum, we must take this sum to at least $k=6$ for fourth-order convergence, $k=8$ for sixth-order, etc.
A similar equation holds for the $v'$ derivative.
Computation of the derivatives of the functions $V_k$, as well as series expansions close to coincidence, is straightforward.
It is also possible to obtain initial data for the higher order and mixed derivatives of the Green function, by simply applying the relevant differential operator to Eq.~\eqref{eq:Ansatz}.

\section{Numerical Results}
\label{sec:Res}

\begin{figure*}[htb!]
	\includegraphics[width = 0.89\columnwidth]{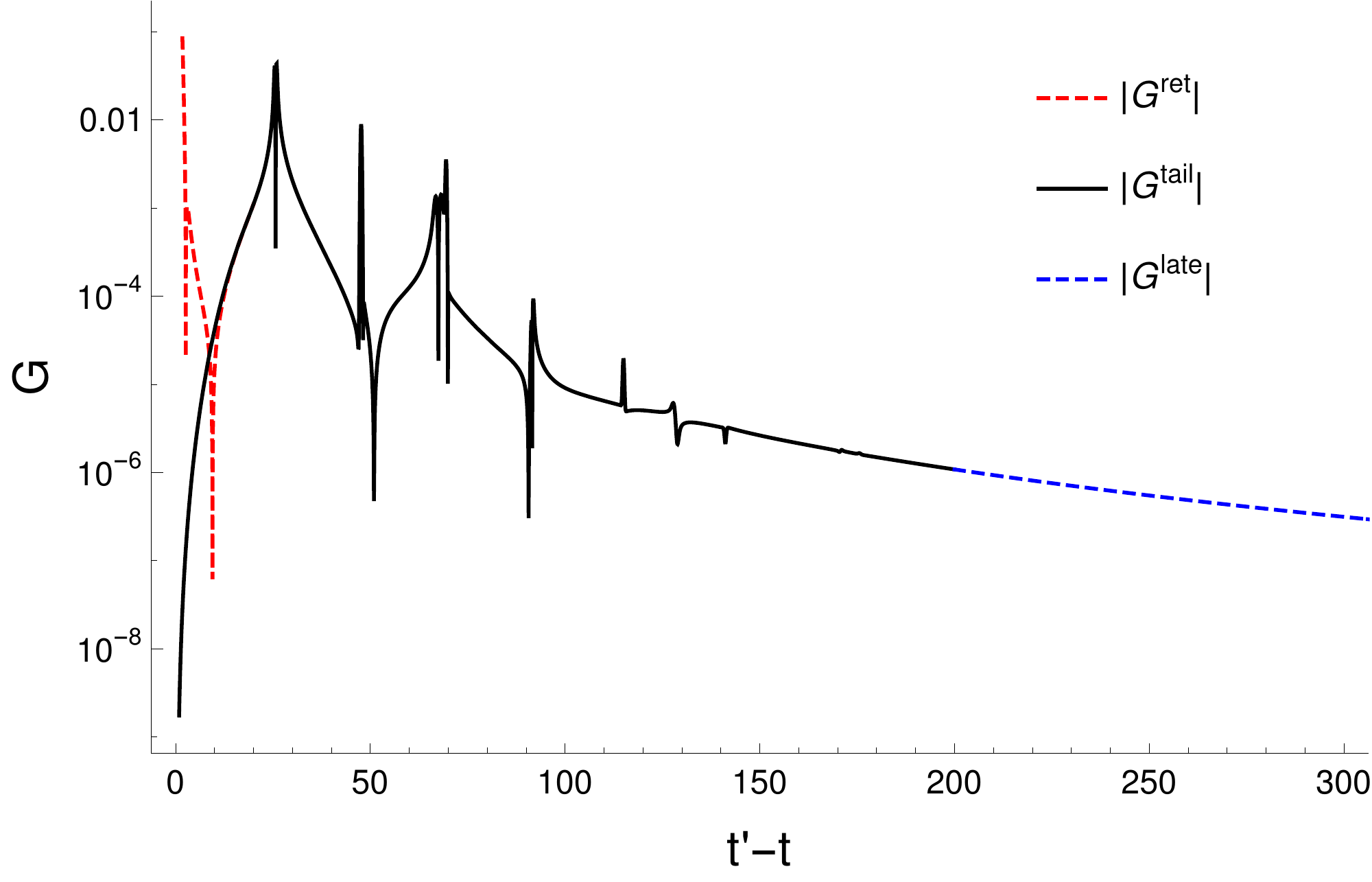}
	\includegraphics[width = 0.89\columnwidth]{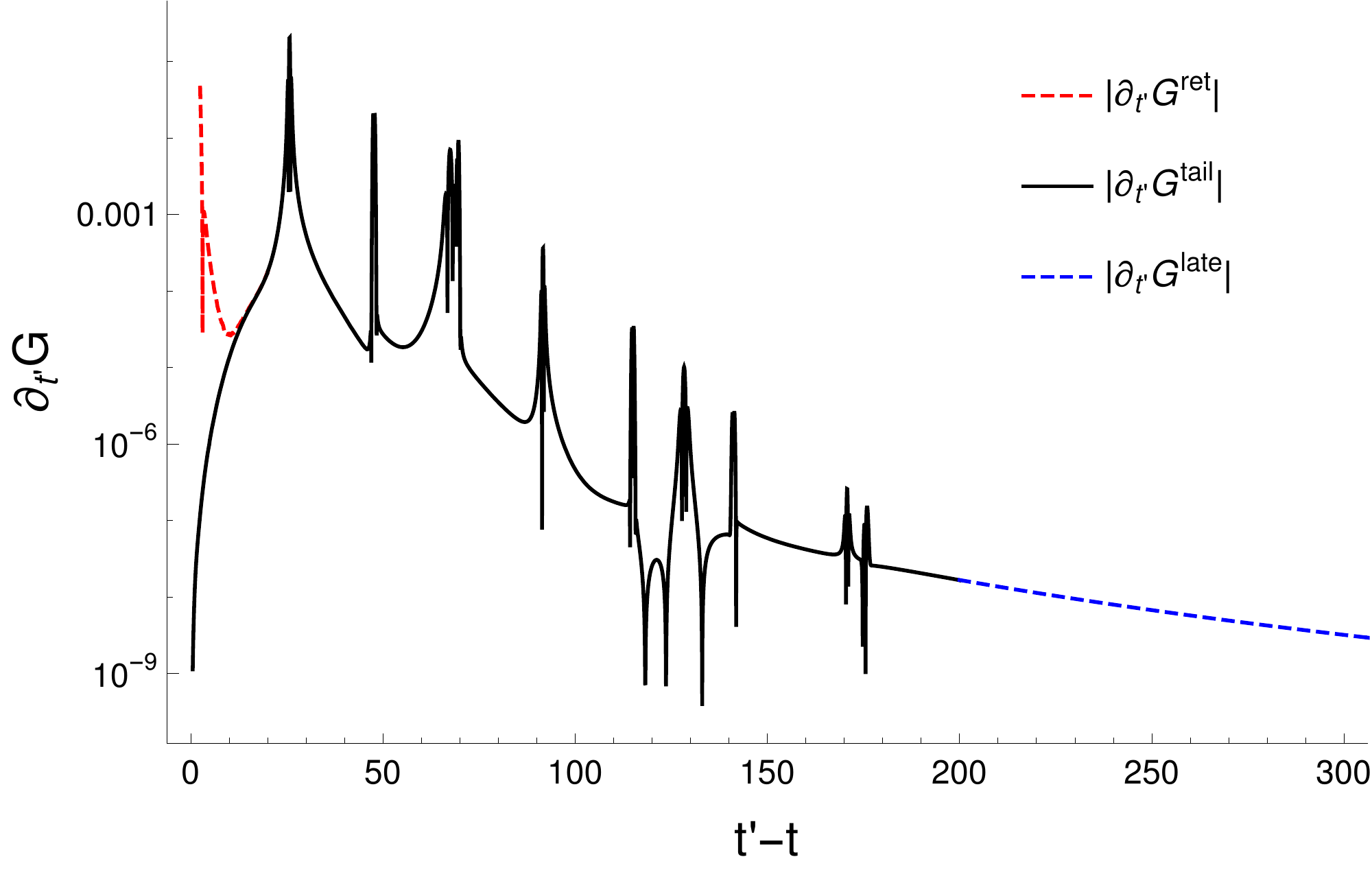}
	\includegraphics[width = 0.89\columnwidth]{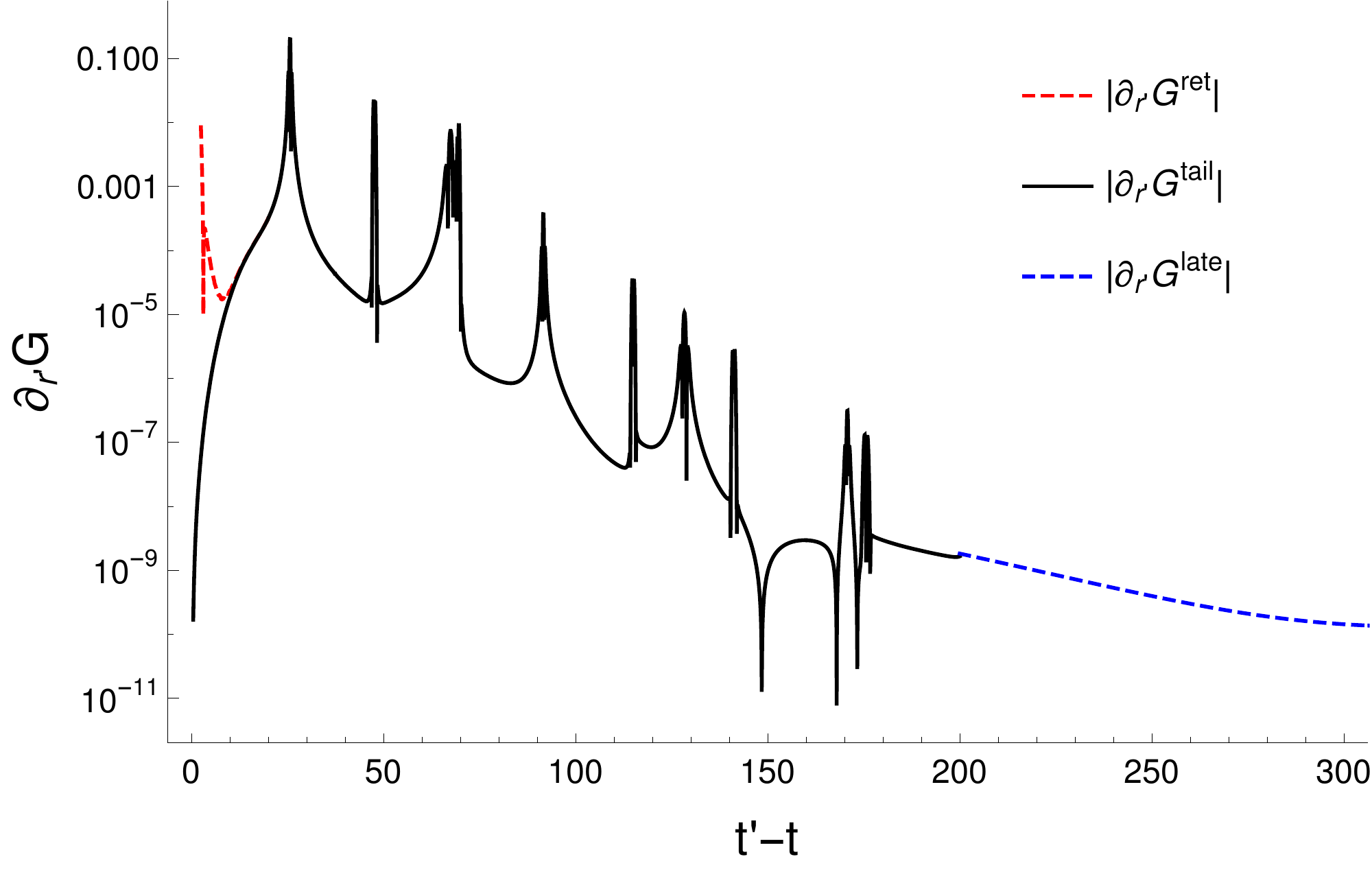}
	\includegraphics[width = 0.89\columnwidth]{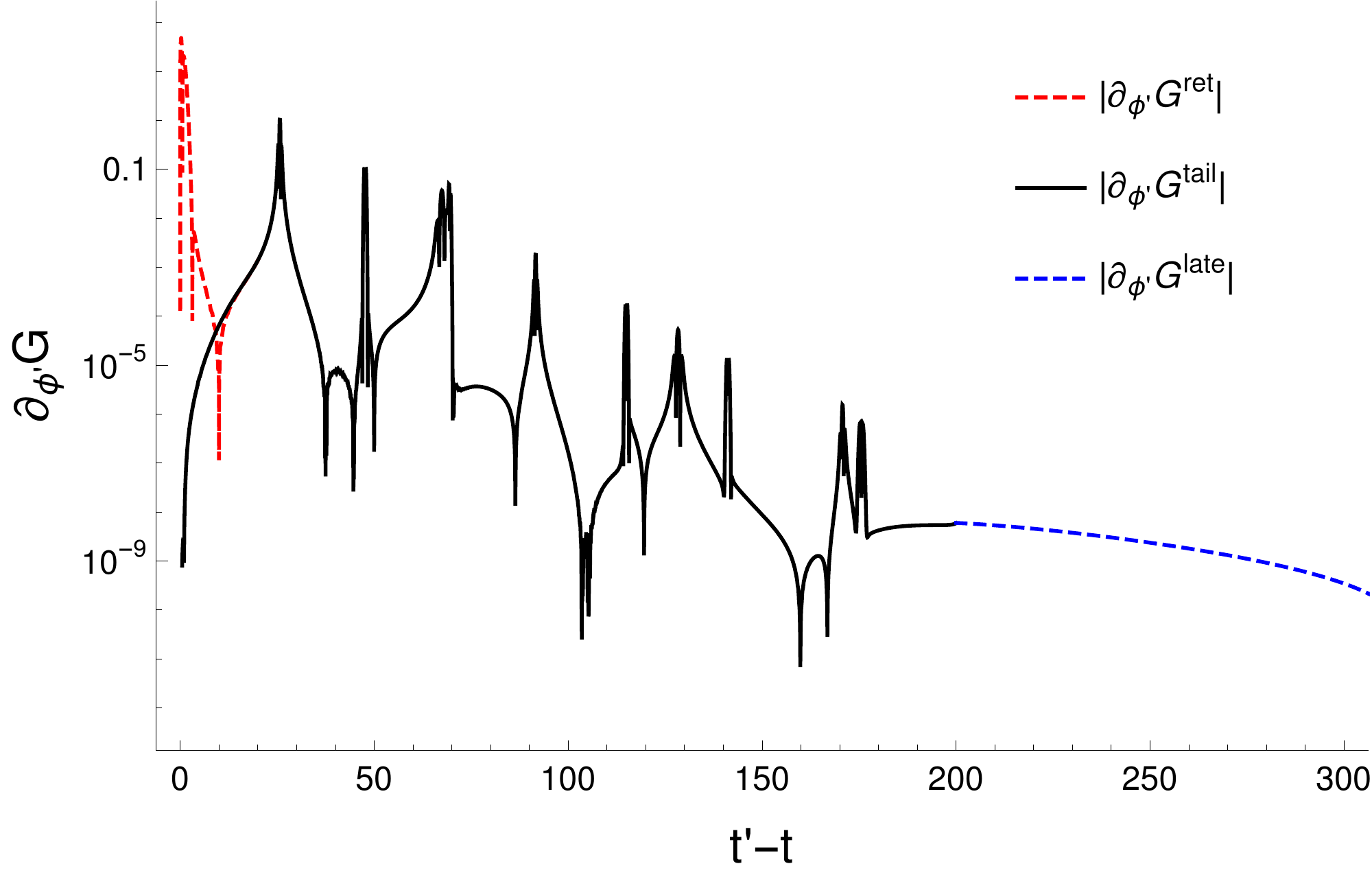}
    \caption[Green function results]{The scalar Green function, $G^{\rm ret}_0(x,x')$ and its derivatives for points separated along an eccentric orbit with $p=7.2$ and $e=0.5$, with base point $r' = 6$.
        The effect of subtracting the direct modes using the method of \cite{Casals:2019heg} to obtain only the tail contribution can be clearly seen in each case. At late times, the numerical solutions also agree with a late-time approximation of the Green function obtained using the method of \cite{Casals:2015nja}.\label{fig:NormRes}}
\end{figure*}

We have implemented the schemes outlined in Sec.~\ref{sec:Num} along with the initial data derived in Sec.~\ref{sec:ICs} as both Mathematica and C codes.
In Fig.~\ref{fig:NormRes} we show some representative Regge-Wheeler results from the C code. These were obtained by
computing the modes $g_{s\ell}$ using our numerical code and following Ref.~\cite{Casals:2013mpa} in performing a smoothed sum over $\ell$,
\begin{equation} \label{eq:resummed}
	G^{\rm ret}_0(x;x') = \frac{1}{r r'}\sum_{\ell=0}^{\ell_{\mathrm{max}}} (2\ell+1) P_{\ell}(\cos\gamma) G^{\rm ret}_{s\ell} (r,r'; \Delta t) e^{-\frac{\ell}{\ell_{\mathrm{cut}}}^2},
\end{equation}
where we chose $\ell_{\rm max}=100$ and $\ell_{\rm cut} = 20$. This smoothing is required to obtain a convergent mode-sum, but
in turn introduces another problem.

To understand this, consider the Hadamard decomposition of the retarded Green function,
\begin{equation}
\label{eq:Hadamard}
	G_0^{\mathrm{ret}}(x;x') = \Theta(x, x')[U(x,x') \delta(\sigma) - V(x,x')\Theta(-\sigma)],
\end{equation}
which is valid in a causal domain of the point $x$ (i.e. where $x$ and $x'$ are connected by a unique geodesic).
Here, the term involving $U(x,x')$ is known as the \emph{direct} part, and the term involving $V(x,x')$ is the \emph{tail} part. As we are constructing the Green
function using a smoothed sum, the $\delta$-distribution in the direct part is effectively smeared into a Gaussian of finite width centred on the base point. This
can clearly be seen in the early-time behaviour of the red dashed curve in Fig.~\ref{fig:NormRes}. To circumvent this, we use the method developed in
\cite{Casals:2019heg}, where it was shown that an $\ell$-mode decomposition of the direct part can be subtracted mode-by-mode from the modes of the full retarded
Green function, $G^{\rm ret}_{s\ell}$. Then, summing over modes we get the full retarded Green function minus its direct part. This regularised Green function
is no longer contaminated by the smeared direct part, as illustrated by the solid black curve in Fig.~\ref{fig:NormRes}.

\begin{figure*}[htb!]
	\includegraphics[width = 0.89\columnwidth]{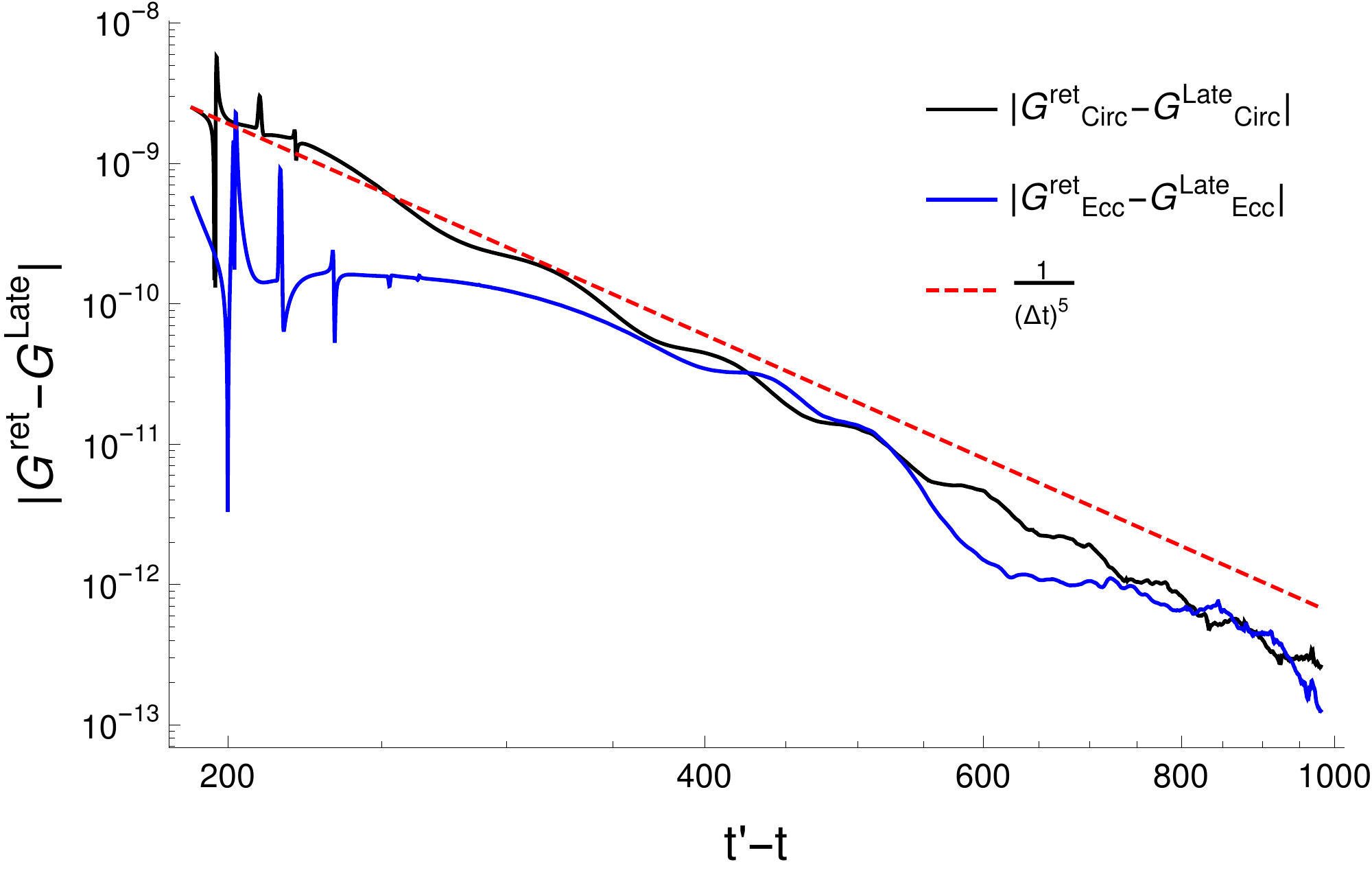}
	\includegraphics[width = 0.89\columnwidth]{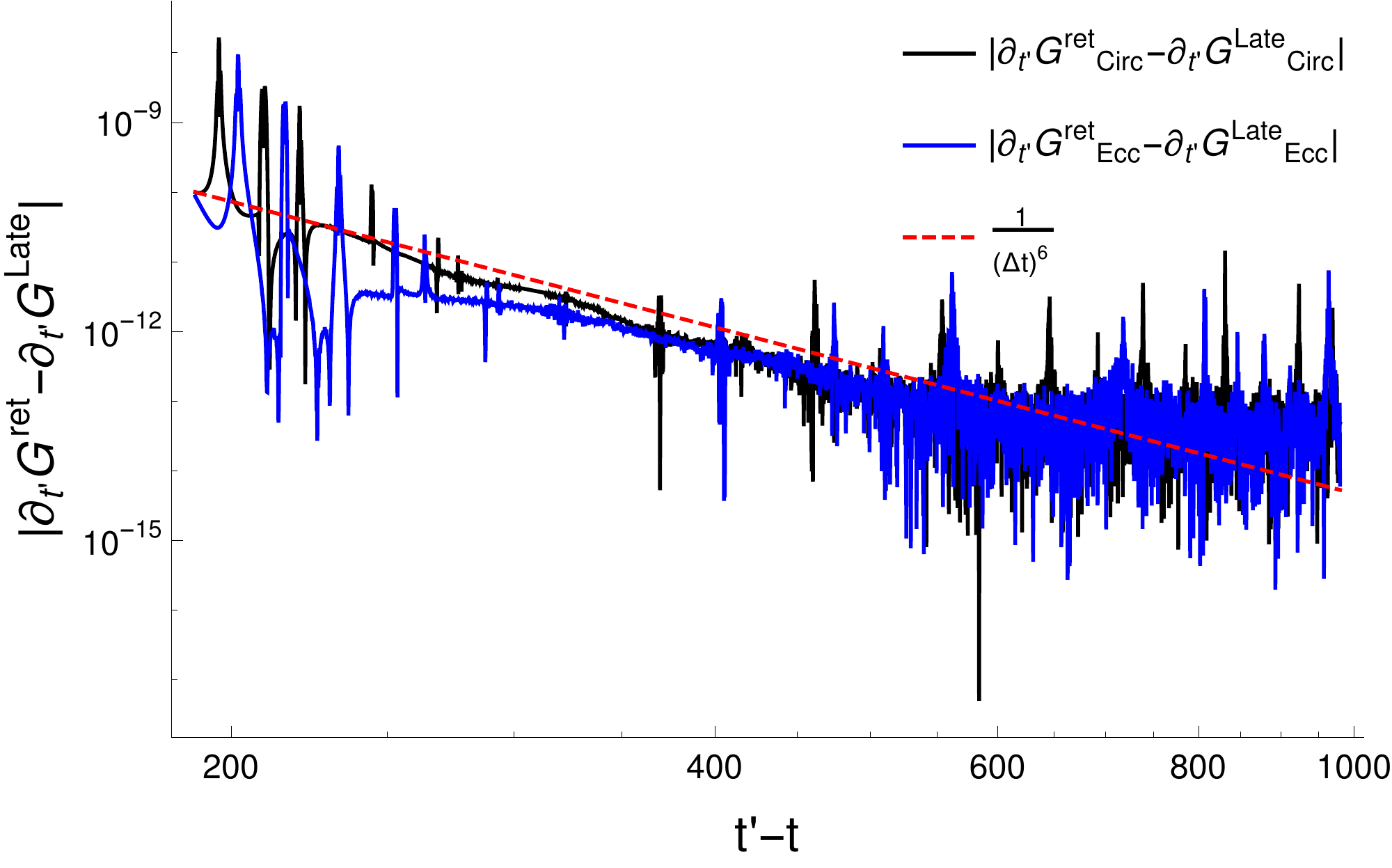}
	\includegraphics[width = 0.89\columnwidth]{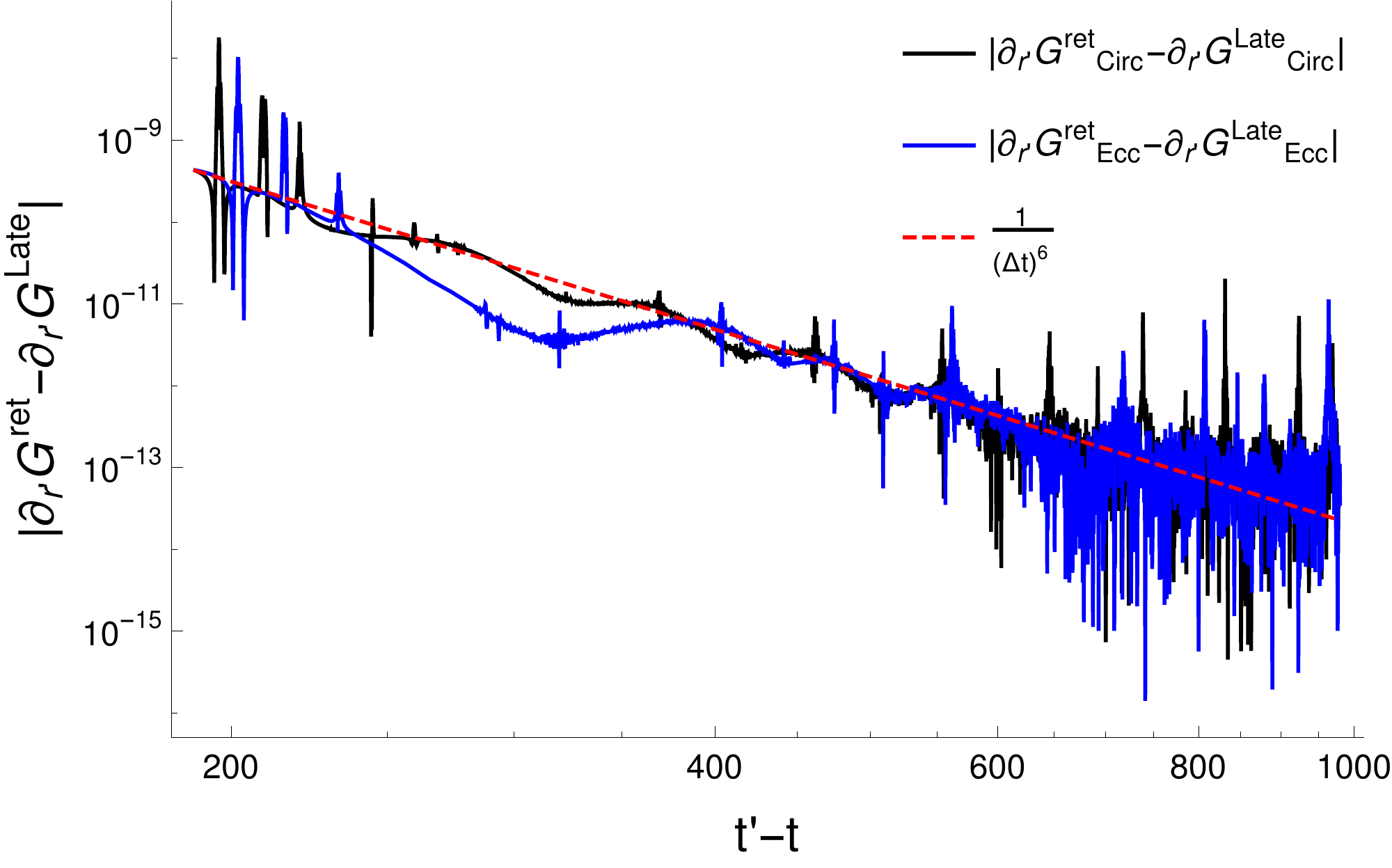}
	\includegraphics[width = 0.89\columnwidth]{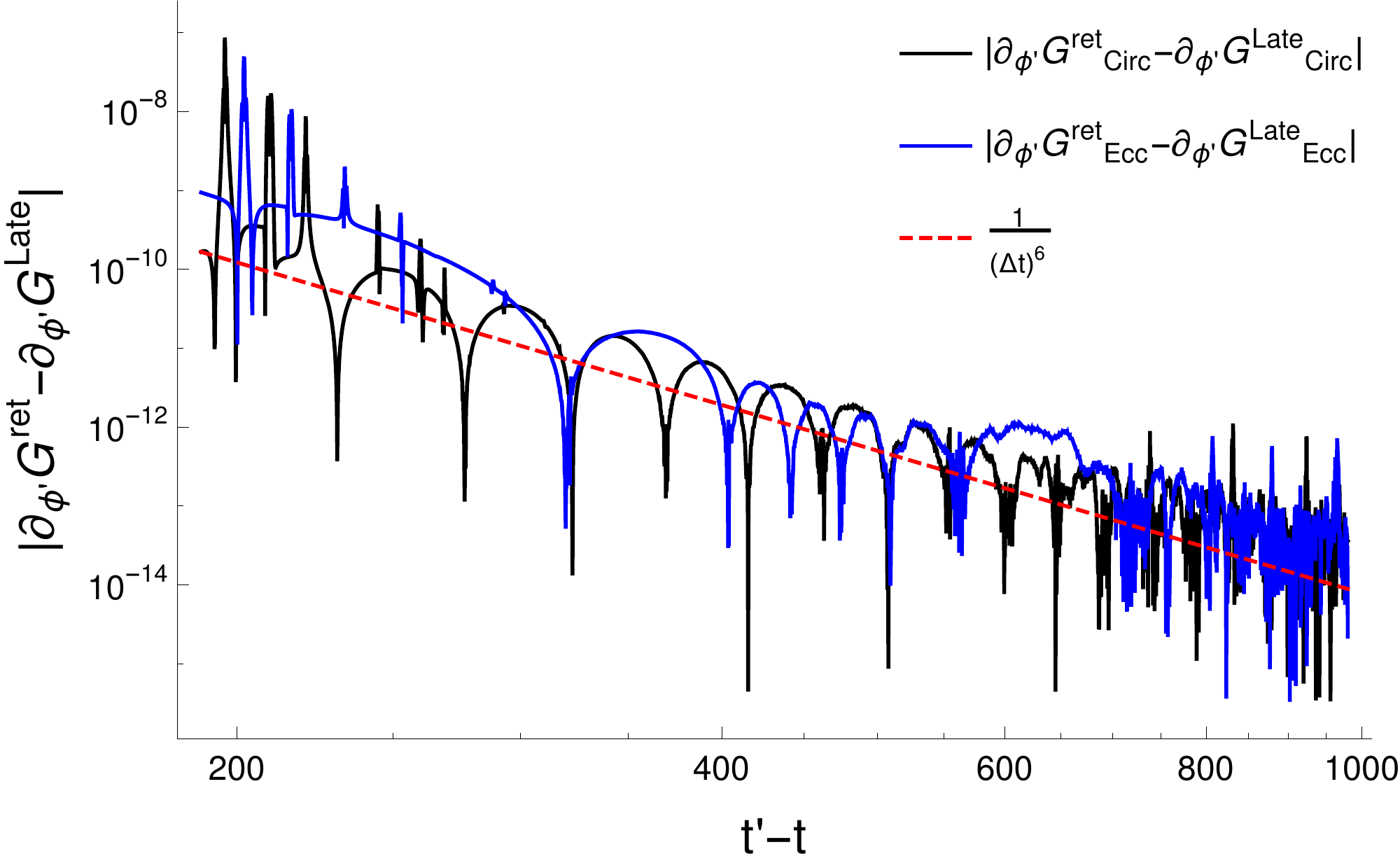}
  \caption[Green function tail match]{The difference between the scalar Green function and its derivatives, and their late-time tail expansions.
        The disagreement behaves as expected given the order at which we truncated the late-time expansions.\label{fig:TailMatch}}
\end{figure*}

\begin{figure*}[htb!]
	\includegraphics[width = 0.9\columnwidth]{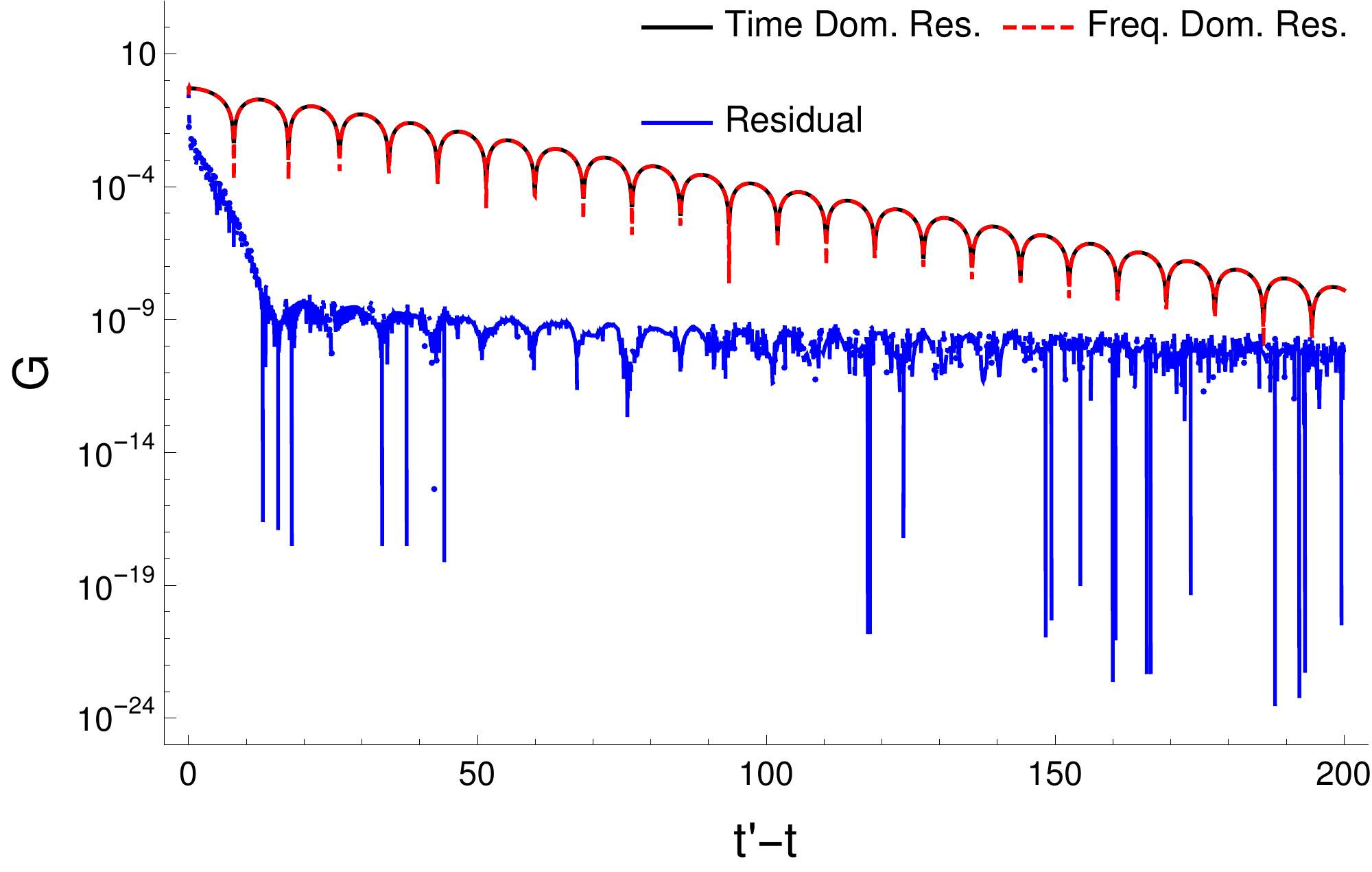}
	\includegraphics[width = 0.9\columnwidth]{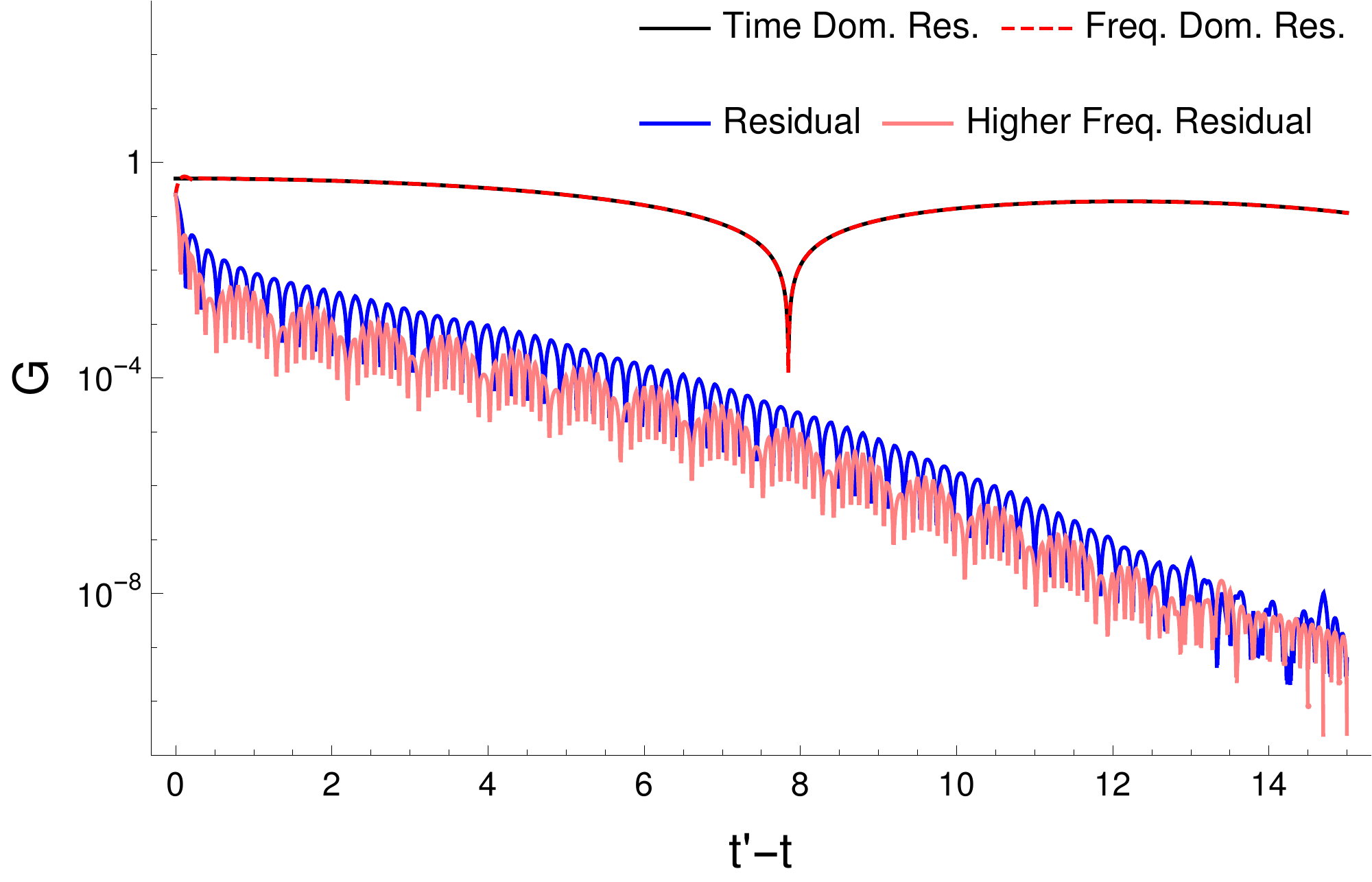}
	\caption[Green function frequency domain vs. time domain]{Comparison of our characteristic time-domain gravitational ($s=2$) Regge-Wheeler Green function against an inverse Fourier transform of
    the frequency domain Green function. The residual depends primarily on the maximum frequency included when inverse Fourier transforming the frequency domain Green function.\label{fig:FreqvsTime}}
\end{figure*}

Our results are consistent with other approaches to computing the Green function, as demonstrated in Figs.~\ref{fig:TailMatch} and \ref{fig:FreqvsTime}, where
we compare with two other approaches to computing it: an expansion valid for late times (Fig.~\ref{fig:TailMatch}) and a numerical calculation based on an
inverse Fourier transform of the frequency-domain Green function (Fig.~\ref{fig:FreqvsTime}). Indeed at sufficiently late times we can save computational cost in Green function calculations by using the late-time expansions in place of our numerically computed solutions.

From Fig.~\ref{fig:NormRes}, it may appear that there is some non-negligible disagreement between the retarded solution and the late-time tail, particularly in the
$r$ derivative of the Green function. However, we can verify that this is consistent with expectations. The leading order term of the late-time tail is $t^{-3}$
and the expansion we use is accurate up to $t^{-5} \log (t)$, taken from \cite{Casals:2015nja}. Thus the late-time tail for the derivatives should be accurate up
to $t^{-6} \log (t)$. We can see in Fig.~\ref{fig:TailMatch}, that as we go to later times, the agreement between the retarded solution and the late-time tail
scales as the next term in the tail expansion. Thus, if greater agreement with the late-time tail is required, it is a simple matter to define a larger numerical
domain and compute the retarded modes to later times. In the self-force calculation described below, however, we found that the relative change from altering
the matching time between the retarded solution and late-time expansion is below our numerical error in the resulting self-field and self-force values.


\subsection{Scalar self-force}

As a demonstration of the utility of our method, we now show how it can be used to compute the scalar self-force.
We obtain results for the self-field and self-forces along both a circular and an eccentric orbit.
Both orbits considered lie close to the separatrix between bound and unbound orbits, with $p = 7.2$, $e=0.5$ in the eccentric case and $p = 6$ in the circular case.

Despite only seeing much of its progress recently \cite{Anderson:2005gb}, the worldline convolution method we use was one of the first methods proposed to tackle the self-force
problem. In two independent works \cite{Mino:1996nk,Quinn:1996am} it was shown that the self-force is given by the so-called MiSaTaQuWa equation, which in the scalar field case and for geodesic motion is given by
\begin{equation}
	\label{eq:MiSaTaQuWa}
	F_{\alpha'}[z(\tau')] = q \lim_{\epsilon \to 0^+}\int_{-\infty}^{\tau'-\epsilon} \nabla_{\alpha'} G_0^{\mathrm{ret}}[z(\tau),z(\tau')] d\tau.
\end{equation}
The integral here is over the past worldline of the compact object, truncated at a proper time just before reaching its current location.
It is clear that due to the cutoff on the upper limit of integration, only the tail part of the Green function contributes to the self-force.
It is also clear that the Hadamard parametrix will be inadequate: the Hadamard decomposition is restricted to a causal domain whereas the integral in
Eq.~\eqref{eq:MiSaTaQuWa} extends over the entire past worldline, which is not confined to a causal domain. Our numerically calculated regularised Green function
does not suffer from this limitation and is therefore ideally suited to computing the self-force using the worldline convolution approach.

With the regularised Green function in hand (and a late-time tail expansion matched on at very late times), we can then compute the self-field and self-force
components by straightforward numerical integration. The results of doing so are shown in Table \ref{tab:Results}. It can be seen that a high degree of accuracy has been achieved.
These results were obtained with $\ell_{\mathrm{max}} = 100$, $\ell_{\mathrm{cut}} = 20$, matching the tail on at $\Delta t=200$, and integrating the tail to $\Delta t=2000$.
We also used a numerical resolution of $h=0.01$ and sampled the numerical solution with resolution of $0.1$ (i.e. every 10 steps).

After performing an analysis on the primary contributions to the numerical error in our results,
the dominant sources were found to be the values of $\ell_{\mathrm{max}}$ and $\ell_{\mathrm{cut}}$.
It is worth noting, however, that we already obtain results of comparable accuracy (particularly in the circular case) to \cite{Wardell:2014kea}, with only half as
many $\ell$-modes used. We attribute this to use of a characteristic initial value formulation, rather than the time-domain Gaussian approach used in \cite{Wardell:2014kea}. This can be verified by replacing our characteristic initial data with a narrow Gaussian of width $0.1$, centred at the base point, and matching on a quasilocal expansion as described in \cite{Casals:2009xa} at $\Delta t=14$, whereby we obtain relative errors of of $\mathcal{O}(1)$\% for the circular orbit self-field when 100 $\ell$-modes are considered.

The derivatives of the direct modes defined in  \cite{Casals:2019heg} are another significant contribution to the error. For convenience, these were computed using finite-differencing (and differentiation of series expansions at early times).
We found that finite differencing error introduced a significant amount of noise if too small a step in $t'$ or $r'$ was chosen or if too high an order of finite differencing was used.
Using larger values of $\Delta t'$ or $\Delta r'$, or a lower order finite difference stencil reduced the noise in the derivatives and lead to changes in the results at $\mathcal{O}(10^{-2})$\%.
Ideally, we would compute these directly rather than via finite difference, but this is beyond the scope of this work.

Another source of error which contributed at a similar order is the sampling resolution.
Increasing the sampling resolution improves the results by $\mathcal{O}(10^{-2})$\%.

The next most significant contributions considered include the time at which the tail is matched on to the numerical results, the maximum time to which the tail is integrated,
and the numerical resolution.
These all contribute at $\mathcal{O}(10^{-3})$\% only.

Finally, the orders of the Bessel series expansions (used for the retarded modes at very early times, as described in \cite{Casals:2019heg}) and the series expansion for the direct modes (again, described in detail in \cite{Casals:2019heg}) contribute negligibly to the error, at $\mathcal{O}(10^{-5})$ and $\mathcal{O}(10^{-8})$\% respectively.

\begin{table}
	\caption{Self-field and self-force components  at $r' = 6$ on two orbits, along with relative errors (as a percentage).\label{tab:Results}}
	\begin{ruledtabular}
		\begin{tabular}{c | c | c | c | c}
			 		       & & Ref. value  & Computed value & Rel. Err. (\%) \\ \hline
			\parbox[t]{2mm}{\multirow{3}{*}{\rotatebox[origin=c]{90}{Eccentric}}} &$\Phi$ & -0.00771731 & -0.00771922    & 0.0247 \\
			&$F_t$      & 0.00066534  & 0.00066633     & 0.1484 \\
			&$F_r$      & 0.00013462  & 0.000134       & 0.459 \\
			&$F_{\phi}$ & -0.00728056 & -0.00728699    & 0.0883 \\ \hline
			\parbox[t]{2mm}{\multirow{3}{*}{\rotatebox[origin=c]{90}{Circular}}} & $\Phi$ & -0.00545483 & -0.00545526 & 0.0079 \\
			&$F_t$      & 0.00036091  & 0.000360997 & 0.0249 \\
			&$F_r$      & 0.00016773  & 0.000167698 & 0.0179 \\
			&$F_{\phi}$ & -0.00530423 & -0.00530738 & 0.0593 \\
		\end{tabular}
	\end{ruledtabular}
\end{table}

\subsection{Gravitational Green function}

In addition to the scalar calculations demonstrated thus far, we can also solve for modes of the Green function and its derivatives in the $s=2$ gravitational case.
While no formal procedure yet exists for subtracting the direct modes or matching on a quasilocal piece at early times, we can still draw comparisons between the results here and frequency domain solutions for the vacuum Regge-Wheeler equation.
Some sample results are show in Fig.~\ref{fig:FreqvsTime}.
These results, for $\ell=2$ along a circular orbit with $r'=6$, show that there is excellent agreement between both the time domain and frequency domain approaches.
While there looks to be larger disagreement at early times, it can be seen in the accompanying plot that increasing the maximum frequency integrated to in the Fourier transformation from
frequency domain to time domain improves the agreement at these early times.
It has also been observed that the magnitude of the residual increases with $\ell$, but again improves with larger maximum frequency.
Thus, the primary source of the residual is the accuracy of the frequency domain results, not the time domain results.

With the accuracy of the gravitational results verified, we next apply the Regge-Wheeler and Zerilli Green functions to the computation of the local force $F_t$ on a gravitational perturbation.
This is directly related to the gravitational energy flux radiated to null infinity and into the horizon, and we use this fact to provide robust reference values against which to check our results.
As demonstration of the method, we choose a circular orbit, $r'=10$, and compute the total flux for specific $(\ell,m)$ modes, $\dot{E}_{21}$, $\dot{E}_{22}$.
This requires no regularisation of the individual $\ell$ modes, and can be constructed easily from the master functions, $\Psi^{\mathrm{RW/Zer}}(t', r_\ast')$, and their derivatives.
In this work, we compute the Moncrief versions of the master functions, detailed in \cite{Martel:2005ir}.

To obtain these, we must integrate the Green function against the source.
The point particle source for the Regge-Wheeler master function, in general, takes the form
\begin{equation}
	S(t', r_\ast') = s_1(t, r_\ast) \delta(r_\ast - r_\ast') + s_2(t, r_\ast) \partial_{r_\ast}\delta(r_\ast - r_\ast').
\end{equation}
Using integration by parts and Eq.~\eqref{eq:Gret}, we obtain an expression for, $\Psi$ at the base point,
\begin{align} \label{eq:Psi} \ \nonumber
&\Psi_{lm}(0, r_\ast')=\pm \frac{1}{2} s_2(0,r_\ast') \\ \nonumber
&\quad + \int_{-\infty}^{0}  \big(g_{2,\ell}(r_\ast', r_\ast';t) \big(\partial_{r_\ast}s_2(t,r_\ast)|_{r_\ast=r_\ast'}-s_1(t,r_\ast')\big) \\
&\quad + \partial_{r_\ast}g_{2,\ell}(r_\ast, r_\ast'';t')|_{r_\ast=r_\ast'} s_2(t,r_\ast') \big) dt
\end{align}
This expression holds for both the Regge-Wheeler-Moncrief and Zerilli-Moncrief master functions, with $g_{2,\ell}$ being the $\ell$ mode of the $s=2$ case of the corresponding Green function.
The $\pm$ comes from the Heaviside step functions in Eq.~\eqref{eq:Gret}, and correctly captures the known jump in the value of the master function at the location of the particle.
Derivatives of $\Psi$ are also required to calculate the flux, but can be easily obtained using the same procedure as for $\Psi$, and computing the derivative before taking the limit to the particle.

The functions $s_1$ and $s_2$, which contain all of the $m$-mode dependence of $\Psi$, can be obtained in terms of the functions $\tilde{F}$, $\tilde{G}$ defined in \cite{Hopper:2010uv}.
Accounting for the authors' convention of fully evaluating the source in terms of $r$ rather than $r_\ast$, expressions for $s_1$ and $s_2$ can be obtained,
\begin{subequations}
	\begin{equation}\label{eq:source}
		s_1(t, r_\ast) =  f^{-1} \tilde{G}(t, r_\ast') + \partial_{r_\ast'}f f^{-2} \tilde{F}(t, r_\ast')
	\end{equation}
	\begin{equation}
		s_2(t, r_\ast) =  f^{-2} \tilde{F}(t, r_\ast').
	\end{equation}
\end{subequations}
When restricted to a circular orbit, the functions $\tilde{F}$ and $\tilde{G}$ are given by
\begin{subequations}
	\begin{equation}
		\tilde{F}^{\mathrm{odd}}_{\ell m} = \frac{8 \pi \mu}{\lambda(\lambda+1)} \frac{f^2 L}{r'} \bar{X}^{\phi}_{\ell m}
	\end{equation}
	\begin{equation}
		\tilde{G}^{\mathrm{odd}}_{\ell m} = -\frac{8 \pi \mu}{\lambda(\lambda+1)} \frac{f L}{r'^2} \bar{X}^{\phi}_{\ell m}
	\end{equation}
	\begin{equation}
		\tilde{F}^{\mathrm{even}}_{\ell m} = \frac{8 \pi \mu}{\Lambda(\lambda+1)} f^2 E  \bar{Y}_{\ell m}
	\end{equation}
	\begin{align}\nonumber\label{eq:Geven}
		\tilde{G}^{\mathrm{even}}_{\ell m} &= -\frac{8 \pi \mu}{\Lambda^2(\lambda+1)} \frac{f E}{r'^3} [\lambda(\lambda+1)r'^2 +6\lambda M r' +15 M^2] \bar{Y}_{\ell m} \\
		& \quad + \frac{8 \pi \mu}{\Lambda(\lambda+1)} \frac{f^3 L^2}{r'^3 E} \bar{Y}_{\ell m} -\frac{8 \pi \mu}{\lambda(\lambda+1)} \frac{f^2 L^2}{r'^3} \bar{Y}^{\phi \phi}_{\ell m}
	\end{align}
\end{subequations}
where $E$ and $L$ denote the energy and angular momentum along the circular orbit, and $Y$, $Y^{\phi \phi}$ and $X^{\phi}$ denote the even sector scalar, tensor
and odd sector vector harmonics, respectively, \cite{Martel:2005ir}.
An overbar denotes complex conjugation.
All instances of $f$ and $\Lambda$ in Eqs.~\eqref{eq:source}--\eqref{eq:Geven} are evaluated at $r'$, rather than $r$.
Note that in this convention the $\partial_{r_\ast}s_2(t,r_\ast)$ term in Eq.~\eqref{eq:Psi} vanishes.

We compute second-order numerical values for the $\ell=2$ mode of the Regge-Wheeler and Zerilli Green functions, as well as all necessary derivatives (note, this computation requires mixed higher order derivatives).
Only the third derivative $\partial_{r'r'r}g_{2,\ell}$ is computed using finite differencing, all others are computed directly.
We thus obtain the flux values shown in Table \ref{tab:Fluxes}.
We expect use of higher order schemes for the Green functions and all relevant derivatives will lead to improved accuracy in the flux calculation.

\begin{table}
	\caption{Gravitational energy fluxes for the $\ell = 2$, $m=1,2$ modes, along with relative errors (as a percentage).
					A circular orbit with $r'=10$ was considered here.
					Reference values obtained using the Black Hole Perturbation Toolkit, \cite{BHPToolkit}.\label{tab:Fluxes}}
	\begin{ruledtabular}
		\begin{tabular}{c | c | c | c | c}
			 		       & & Ref. value  & Computed value & Rel. Err. (\%) \\ \hline
			&$\dot{E}_{21}$      & 9.719$\times 10^{-8}$  & 9.833$\times 10^{-8}$     & 1.17 \\
			&$\dot{E}_{22}$ & 2.685$\times 10^{-5}$ & 2.753$\times 10^{-5}$    & 2.54 \\
		\end{tabular}
	\end{ruledtabular}
\end{table}

\section{Conclusions\label{sec:Conc}}

In this work, we have implemented and demonstrated an efficient means of computing the retarded Green function for the Regge-Wheeler and Zerilli equations.
We derive, for the first time, initial data \emph{inside} the lightcone to arbitrary order in the distance from the lightcone. This initial data can be used to seed standard numerical approaches.

We also implement methods of removing the direct part of the scalar Green function and its derivatives for the first time,
resulting in the computation of accurate self-field and self-force values for both a circular and eccentric orbit. In future works, this method can be trivially applied to compute the self-force along more complex worldlines, such as a hyperbolic encounter with the central black hole.

The improvements over previous methods are apparent, with the number of modes required approximately halved, without a significant loss of accuracy.
The numerical scheme and initial data may also be systematically extended to any numerical order in the grid-spacing $h$, allowing for potentially further gains in efficiency, via coarser grids which yield the same accuracy.
Additionally, a large portion of the calculations detailed here were performed in Mathematica.
Converting these to C may take time, and require the use of third-party libraries such as the GNU Scientific Library.
The gain in computational speed, however, would make this a worthwhile effort.

Currently, the greatest outstanding issue is the means of removing the direct part of the Green function in the gravitational case.
With this done, the results could be applied to compute  full metric perturbations in the Regge-Wheeler gauge, and gauge invariants used to draw comparison with other approaches to the two-body problem.

Finally, many of the methods developed in this work can be applied to other problems, such as the Lorenz gauge wave equation for metric perturbations of Schwarzschild spacetime, or even the the Teukolsky equation. The efficiency of the code also makes other numerically intense applications more feasible, such as a self-consistent evolution scheme.

\begin{acknowledgments}

The authors thank Marc Casals, David Aruquipa, Brien Nolan and Leor Barack for helpful discussions, as well as Josh Mathews for providing reference values for the flux.
COT acknowledges support from the Irish Research Council under grant GOIPG/2017/1031.
This work makes use of the Black Hole Perturbation Toolkit.

\end{acknowledgments}

\bibliography{RWGreenFunc}

\end{document}